\documentclass[aps,pra,twocolumn,showpacs,preprintnumbers]{revtex4}
\usepackage{dcolumn}
\usepackage{bm}
\usepackage{graphicx}
\usepackage{amsmath}
\usepackage{latexsym}
\usepackage{txfonts}
\usepackage{amsfonts}
\usepackage{amssymb}
\usepackage{array}
\usepackage{epsfig}
\usepackage{mathrsfs}
\allowdisplaybreaks[1]
\usepackage{epstopdf}

\newcommand{\ket}[1]{\left\vert#1\right\rangle}

\newcommand{\be}{\begin{equation}}
\newcommand{\ee}{\end{equation}}
\newcommand{\ba}{\begin{array}}
\newcommand{\ea}{\end{array}}
\newcommand{\bea}{\begin{eqnarray}}
\newcommand{\eea}{\end{eqnarray}}

\DeclareSymbolFont{symbols}{OMS}{cmsy}{m}{n}

\begin{document}
\title{Quantum theory of surface-plasmon polariton scattering}

\author{D. Ballester,$^1$ M. S. Tame,$^{1,2,3}$ and M. S. Kim$^{1,2,3}$}
\affiliation{$^1$School of Mathematics and Physics, Queen's University,~Belfast BT7 1NN, United Kingdom \\$^2$ Institute for Mathematical Sciences, Imperial College London, SW7 2PG, United Kingdom \\ $^3$QOLS, The Blackett Laboratory, Imperial College London, Prince Consort Road, SW7 2BW, United Kingdom}

\date{\today}
   
\begin{abstract} 
We introduce the quantum mechanical formalism for treating surface plasmon polariton scattering at an interface. Our developed theory - which is fundamentally different from the analogous photonic scenario - is used to investigate the possibility of plasmonic beamsplitters at the quantum level. Remarkably, we find that a wide-range of splitting ratios can be reached. As an application, we characterize a 50:50 plasmonic beamsplitter and investigate first-order quantum interference of surface plasmon polaritons. The results of this theoretical study show that surface plasmon beamsplitters are able to reliably and efficiently operate in the quantum domain.
\end{abstract}

\pacs{03.67.-a, 73.20.Mf, 42.50.Ex, 03.70.+k}
\maketitle

\section{Introduction}
Nanophotonic systems based on surface plasmon polaritons (SPPs)~\cite{Zayats,photoncircuit} are currently raising considerable interest from the quantum optics and quantum information communities~\cite{Alte,plasmonQIP,Lukin1,Lukin2,TSPP,DSPP,Lukin3,Kol}. Due to their tight-field confinement \cite{Gramotnev} and 
electro-optical behavior~\cite{Zayats,photoncircuit,Maier2}, SPPs constitute compact and versatile candidates for quantum information processing (QIP) with light at the nanoscale. One of the most important ingredients for plasmonic-based QIP is the ability for SPPs to interact coherently with each other. Recent theoretical~\cite{Oult,Elser} and experimental work~\cite{Ditl,Hon,Ebbes,Boz} has hinted at the possibility of achieving coherent interactions between SPPs via scattering type processes. Here, metal-dielectric interfaces~\cite{Steg,Oult,Elser}, as well as junctions and splitters in waveguides structures~\cite{Ebbes,Boz} have been considered. However, this work has so far been restricted to a purely classical domain and little is known about surface plasmon interactions at the quantum level. The development of a flexible theory that can be applied to a variety of different types of waveguide geometries should greatly aid the design of plasmonic components exploiting quantum mechanical effects and circuitry for QIP applications.

In this work we introduce and develop the quantum mechanical formalism for treating SPP interactions via scattering at an interface. Our theory is then used to investigate the possibility of constructing plasmonic beamsplitters that are able to operate faithfully and efficiently at the quantum level. Surprisingly, we find that a wide-range of splitting ratios can be reached without the need for embedding complex optical material, such as anisotropic metamaterials, as recently suggested~\cite{Elser}. We also find that power loss due to unavoidable so-called `parasitic scattering' of SPPs into photon radiation~\cite{Oult} can be suppressed to $5\%$ or even less in some cases. Moreover, the beamsplitter geometries we investigate are directly accessible to experiments; due to the basic properties of the materials involved, complicated on-chip fabrication techniques are not required. As an application of our theory - and as an example of the necessity for a quantum theory of scattering - we optimize a 50:50 plasmonic beamsplitter and use it to investigate first-order quantum interference effects of SPPs. Our study shows that surface plasmon beamsplitters can reliably operate at the quantum level, providing important insights and helping to open up new directions of research into the design of efficient and practical components for on-chip plasmonic-based QIP.

The paper is structured as follows: In Section II we introduce the interface scenario considered for SPP interactions via scattering. Here, a quantized description of all the fields involved is presented and a brief discussion of relevant material properties is included. In Sections III and IV we formalize our theory for the scattering process, where we provide a normalization procedure for the quantized fields and introduce field-matching relations. This enables the formation of a quantum transfer matrix in Section V linking all the relevant excitations together. In Section VI we use our theory to investigate the possibility of efficient quantum plasmonic beamsplitters. Here, we discuss important issues such as spatio-temporal indistinguishability, reciprocity and loss effects. In Section VII we use the results from the previous sections to investigate first-order quantum interference of SPPs. Finally, Section VIII summarizes our main results.

\section{Interface Configuration}
\begin{figure*}[t]
\centerline{\psfig{figure=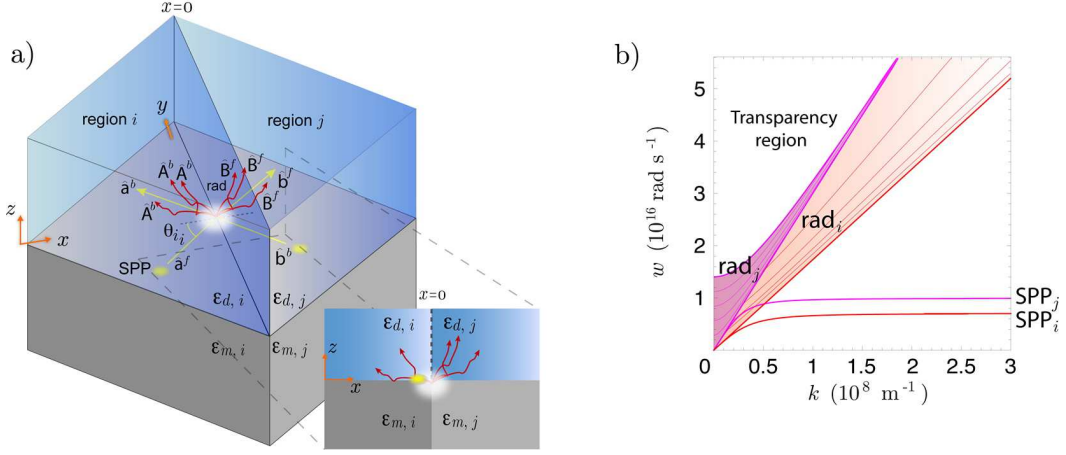,width=14cm}}
\caption{(Color online) Scattering of SPPs at an interface. 
{\bf (a)}: An SPP excitation $\hat{a}^f$ in region $i$ is scattered into an SPP excitation $\hat{b}^f$ in region $j$ and SPP excitation $\hat{a}^b$ in region $i$. The superscript $f$ ($b$) denotes a forward (backward) propagating excitation with respect to the $x$ axis. Near- and far-field radiation excitations are also excited in the process (represented by the central white haze and red jagged arrows). The symbols $\hat{A}^b_k$ and $\hat{B}^f_k$ are used for the radiation in regions $i$ and $j$ respectively. Similar scattering occurs for the SPP excitation $\hat{b}^b$ in region $j$. Further details on the angles are given in Fig. 2 {\bf (b)} and related discussion in Section V. {\bf (b)}: Dispersion relations for the SPP's and radiation in regions $i$ and $j$. Shaded regions correspond to a continuum spanned by the wavenumber $q_i$ and $q_j$ respectively. Examples of lines corresponding to set $q_i$ and $q_j$ are shown.}
\label{fig1}
\end{figure*}

The interface considered is shown in Fig.~\ref{fig1}~{\bf (a)}. Here two regions $i$ and $j$, with different materials are combined at $x=0$ (inset shows the cross-section). Region $i$~($j$) consists of metal with permittivity $\epsilon_{m,i}$ ($\epsilon_{m,j}$) for $z<0$ and a dielectric media with permittivity $\epsilon_{d,i}$ ($\epsilon_{d,j}$) for $z\ge0$. Before we introduce the quantum formalism, a short description of the system dynamics is given. The purpose of this is to provide an informative glimpse of the detailed theory that is to follow. 

In Fig.~\ref{fig1}~{\bf (a)} an SPP excitation, denoted by the operator $\hat{a}^f$, is shown in region $i$ moving forward from the left at an angle $\theta_{i_i}$. This SPP scatters at the interface ($x=0$) into a backward moving SPP excitation, denoted by the operator $\hat{a}^b$, in region $i$ at an angle $\theta_{i_r}$, in addition to a forward moving SPP excitation, denoted by the operator $\hat{b}^f$, in region $j$ at an angle $\theta_{i_t}$, as well as into near- and far-field photon radiation excitations moving backwards and forwards in regions $i$ and $j$ respectively, denoted by the operators $\hat{A}^b$ and $\hat{B}^f$ (all angles are in the $x$-$y$ plane and shown in more detail in Fig.~\ref{fig2}~{\bf (b)}). The radiation excitations, have a range of in-plane and out-of-plane wavevector components (with respect to the $x$-$y$ plane) and act as a loss mechanism for SPP interactions at the interface. Later, we will show how this loss can be suppressed to less than $5\%$ by carefully modifying the properties of the metal and media in regions $i$ and $j$, as well as the incidence angle of the incoming SPP. A similar outline to the above can also be given for an SPP excitation moving from region $j$ to $i$.

We now formalize the above scenario. In order to describe the scattering process quantum mechanically, we start by introducing the quantized vector potential field operators for the various excitations supported by the geometry of Fig.~\ref{fig1}~{\bf (a)} and involved in the interaction. We consider the vector potential as it provides a convenient starting point from which to derive both the quantized electic and magnetic field operators~\cite{Loudon}. These operators will play a central role in our theory.

\subsection{Quantized radiation field}
In region $i$ both transverse magnetic (TM) and transverse electric (TE) surface radiation fields are supported~\cite{ER,SnyderLove}. This is in contrast to SPPs which are restricted to TM fields, as we will discuss in the next section.
The quantized vector potential for the radiation fields is derived from the field equation for the geometry in the Coulomb gauge with classical-quantum correspondence relations applied~\cite{ER}. It is given by 
\bea
&&\hskip-0.50cm\hat{\bf A}_{ri}({\bf r},t)=\frac{1}{2 \pi} \int {\rm d}{{\bf
k}_i}\int_{0}^{q_{\rm cut}} {\rm d}q_i \left(\frac{\hbar c^2 q_i^2 }{\epsilon_0 \pi \omega_i^3 }\right)^{1/2} \times \label{vecr} \\
&&\hskip-0.15cm[{\bm \phi}_r({{{\bf
k}_i},q_i},{\bf
r})\hat{A}_r({\bf
k}_i,q_i)e^{-i \omega_i t}+ {\bm \phi}_r^*({{{\bf
k}_i},q_i},{\bf
r})\hat{A}^\dag_r({\bf
k}_i,q_i)e^{i \omega_i t}]. \nonumber
\eea
The creation (annihilation) operators $\hat{A}^\dag_r({\bf
k}_i,q_i)$ ($\hat{A}_r({\bf
k}_i,q_i)$), with $r\in\{{\textit {\footnotesize TM}},{\textit {\footnotesize TE}}\}$, satisfy the bosonic commutation relations $[\hat{A}_r({\bf
k}_i,q_i),\hat{A}_{r'}^\dag({\bf
k}_i',q_i')]=\delta({\bf
k}_i-{\bf
k}_i')\delta(q_i-q_i')\delta_{rr'}$ from which the Heisenberg uncertainty relations are derived~\cite{Loudon}. The wavefunctions are given by
\begin{eqnarray}
&&\hskip-0.5cm{\bm \phi}_{{\textit {\tiny TM}}}({\bf k}_i,q_i,{\bf
r})=e^{i {\bf k}_i \cdot {\bf r}}  \gamma_i^{{\textit {\tiny TM}}}  \bigg[ (i \hat{{\bf k}}_i+\frac{k_i}{\nu_i}\hat{{\bf z}})e^{\nu_i z} \vartheta (-z)   \,+ \label{modrTM} \\
&&\hskip-0.0cm\big[i \hat{{\bf k}}_i(\cos q_i z - \eta_i \sin q_i z)+\frac{k_i}{q_i}\hat{{\bf z}}(\sin q_i z + \eta_i \cos q_i z)\big] \vartheta (z) \bigg] \nonumber
\end{eqnarray}
and
\begin{eqnarray}
&&\hskip-0.5cm{\bm \phi}_{{\textit {\tiny TE}}}({\bf k}_i,q_i,{\bf
r})=e^{i {\bf k}_i \cdot {\bf r}}  \gamma_i^{{\textit {\tiny TE}}}  i (\hat{{\bf z}} \times \hat{{\bf k}}_i)[e^{\nu_i z} \vartheta (-z)   \,+ \label{modrTE} \\
&&\hskip3.5cm(\cos q_i z + \frac{\nu_i}{q_i} \sin q_i z) \vartheta (z)], \nonumber
\end{eqnarray}
with $\vartheta (z)$ representing the heaviside function,
${\bf k}_i=k_{x,i} \hat{{\bf x}}+k_{y,i} \hat{{\bf y}}$ is a wavevector with wavenumber $k_i=(k_{x,i}^2+k_{y,i}^2)^{1/2}$, $q_i=(\omega_i^2 \epsilon_{d,i}/c^2 -k^2_i)^{1/2}$ characterizes the variation of the wavefunctions in the region $z \ge 0$ and $\nu_i=(k^2_i-\omega_i^2 \epsilon_{m,i} / c^2)^{1/2}$ characterizes the $z < 0$ variation. In addition, $\epsilon_{d,i}$ is the dielectric function of the non-metallic media in region $i$ and $\epsilon_{m,i}=1-\omega_{p_i}^2/\omega_i^2$ is the dielectric function of the metal, with $\omega_{p_i}$ as the plasma frequency in region $i$ and the parameters $\eta_i=q_i \epsilon_{m,i}/\nu_i \epsilon_{d,i}$, $\gamma_i^{{\textit {\tiny TM}}} = \nu_i/(\epsilon_{d,i}^2 \nu_i^2 + \epsilon_{m,i}^2 q_i^2 )^{1/2} $ and $\gamma_i^{{\textit {\tiny TE}}} =[\epsilon_{d,i}(\epsilon_{d,i}-\epsilon_{m,i})]^{-1/2}$~\cite{ER,Oultrec}. Here, the Drude model~\cite{Zayats} is used in $\epsilon_{m,i}$ for the purpose of illustrating the results of our investigation and we have extended the work of Ref.~\cite{ER} to allow for an arbitrary dielectric function $\epsilon_{d,i}$ that is real and positive. 
The parameter $q_i$ is chosen to be the free parameter characterizing the complete $z$ variation of the wavefunctions. This can be done by substituting $k_i$ in terms of $q_i$ in the definition of $\nu_i$ giving $\nu_i=(\omega_i^2 (\epsilon_{d,i}-\epsilon_{m,i})/ c^2-q_i^2)^{1/2}$. Thus the definition of $q_i$ provides a dispersion relation (DR) which links $\omega_i$ to $k_i$ regardless of the polarization (TM or TE). This DR is set by the value of $q_i$, once $\epsilon_{d,i}$ is specified. In Fig.~\ref{fig1}~{\bf (b)} we show example DRs of the radiation in separate regions $i$ and $j$ for $\epsilon_{d,i}=3$ and $\epsilon_{d,j}=1$. For $q=0$, the radiation has the usual bulk photon DR expected from a medium with dielectric function $\epsilon_d$, given by $k=\sqrt{\epsilon_d}\omega/c$. However, as $q$ increases, the penetration of the field into the metal, characterized by the $e^{\nu z}$ term in Eqs.~(\ref{modrTM})~and~(\ref{modrTE}), increases and the DR is modified substantially~\cite{ER}. The shaded regions denoted by rad$_i$ and rad$_j$ correspond to a continuum of surface photon DRs spanned by $q_i$ and $q_j$ respectively. 

We limit our work to frequencies below $\omega_{p,i}$. Here the integral over $q_i$ in Eq.~(\ref{vecr}) explicitly covers the range 0 to a maximum cutoff value $q_{\rm cut}=\omega_i (\epsilon_{d,i})^{1/2}/c$. For $q_i>q_{\rm cut}$, $k_i$ becomes imaginary and for $q_i>\omega_i(\epsilon_{d,i}-\epsilon_{m,i})^{1/2}/c>q_{\rm cut}$, $k_i$ and $\nu_i$ are imaginary. Both these ranges require a different analytical form for $\hat{\bf A}_{r i}({\bf r},t)$. For frequencies above $\omega_{p,i}$ we have $q_{\rm cut}>\frac{\omega_i}{c}(\epsilon_{d,i}-\epsilon_{m,i})^{1/2}$ and the regime $\frac{\omega_i}{c}(\epsilon_{d,i}-\epsilon_{m,i})^{1/2}<q_i<q_{\rm cut}$ (where $k_i$ is real but $\nu_i$ is imaginary) corresponds to the transparency region of the metal, as shown in Fig.~\ref{fig1}~{\bf (b)}. 
The quantized vector potential for the radiation field in region $j$ is given by Eqs.~(\ref{vecr}),~(\ref{modrTM})~and~(\ref{modrTE}) with the relabeling $i \to j$ and the creation (annihilation) operators $\hat{A}_r^\dag \to \hat{B}_r^\dag$ ($\hat{A}_r \to \hat{B}_r$).

\subsection{Quantized SPP field}
The quantized vector potential for the SPP field in region $i$ is derived in a similar way to the radiation field~\cite{ER,TSPP,DSPP,SPPQ}. However unlike the radiation field, due to the materials and geometry considered here, SPP boundary conditions lead only to TM fields being supported~\cite{Zayats,Oult,Elser}. The quantized vector potential is given by~\cite{ER}
\bea
&&\hskip-0.45cm\hat{\bf A}_{pi}({\bf r},t)=\frac{1}{2 \pi} \int {\rm d}{\bf
k}_i\left(\frac{\hbar}{2 \epsilon_0 \omega_i p_i}\right)^{1/2} \times \label{vecp} \\
&&\hskip1.8cm[{\bm \phi}_p({\bf
k}_i,{\bf
r})\hat{a}({\bf
k}_i)e^{-i \omega_i t}+ {\bm \phi}_p^{*}({\bf
k}_i,{\bf
r})\hat{a}^\dag({\bf
k}_i)e^{i \omega_i t}],\nonumber
\eea
with creation (annihilation) operators $\hat{a}^\dag({\bf
k}_i)$ ($\hat{a}({\bf
k}_i)$) satisfying the bosonic commutation relations $[\hat{a}({\bf
k}_i),\hat{a}^\dag({\bf
k}_i)]=\delta({\bf
k}_i-{\bf
k}_i')$ and wavefunctions 
\bea
\hskip-0.25cm{\bm \phi}_p({\bf k}_i,{\bf
r})&=&\big[(i \hat{{\bf k}}_i+\frac{k_i}{ \nu_i}\hat{{\bf z}})e^{\nu_i z} \vartheta (-z)+ \label{modp} \\
&& \hskip1.0cm\qquad (i \hat{{\bf k}}_i-\frac{k_i}{ \nu_{0,i}}\hat{{\bf z}})e^{-\nu_{0,i} z} \vartheta (z) \big] e^{i {\bf k}_i \cdot {\bf r}}. \nonumber
\eea
Here, ${\bf k}_i=k_{x,i} \hat{{\bf x}}+k_{y,i} \hat{{\bf y}}$ is again the wavevector with wavenumber $k_i=(k_{x,i}^2+k_{y,i}^2)^{1/2}$, 
$\nu_{0,i}=(k^2_i-\omega_i^2 \epsilon_{d,i}/c^2)^{1/2}$ characterizes the variation of the wavefunctions in the region $z \ge 0$ and $\nu_i=(k^2_i-\omega_i^2 \epsilon_{m,i} / c^2)^{1/2}$ characterizes the $z<0$ variation. 
The parameter $p_i=ct_i[2\omega_i\epsilon_{m,i}^2\epsilon_{d,i}(-(\epsilon_{m,i}+\epsilon_{d,i}))^{1/2}]^{-1}$, with $t_i=(\epsilon_{m,i}^2+\epsilon_{d,i})(\epsilon_{m,i}^2-\epsilon_{d,i}^2)$. Again, here we have extended the work in Ref.~\cite{ER} in order to allow for an arbitrary real and positive $\epsilon_{d,i}$.
A DR that is independent from $\nu_{0,i}$ and $\nu_{i}$ can be derived from boundary conditions and is given by $k_i=(\omega_i/c)(\epsilon_{d,i}\epsilon_{m,i}/(\epsilon_{d,i}+\epsilon_{m,i}))^{1/2}$~\cite{ER,TSPP,DSPP,SPPQ}. Once $k_i$ is set, with $\epsilon_{d,i}$ and $\epsilon_{m,i}$ both specified, $\omega_i$ is set and so are $\nu_i$ and $\nu_{0,i}$ by definition. This is in direct contrast to the radiation excitations, where for a set wavenumber $k_i$ there are a range of values of $\omega_i$ which satisfy the DR, dependent on the free parameter $q_i$. Here, as $\nu_i$ and $\nu_{0,i}$ are both set by $k_i$, for SPPs there is no longer a free parameter for the $z$ variation. Thus neither $\nu_i$ nor $\nu_{0,i}$ appears explicitly as an argument of the wavefunctions. In Fig.~\ref{fig1}~{\bf (b)} we show example DRs of SPPs in regions $i$ and $j$ for $\epsilon_{d,i}=3$ and $\epsilon_{d,j}=1$, with $\epsilon_{m}$ chosen to be that of silver, $\omega_{p,i}=\omega_{p,j}=1.402\times 10^{16}~{\rm rad~s}^{-1}$~\cite{JohnChrist}. The quantized vector potential for the SPP field in region $j$ is given by Eqs.~(\ref{vecp})~and~(\ref{modp}) with the relabeling $i \to j$ and the creation (annihilation) operators $\hat{a}^\dag \to \hat{b}^\dag$ ($\hat{a} \to \hat{b}$).

By inspection, the wavefunctions for the radiation and SPP excitations for a given region $i$ are found to be orthogonal~\cite{ER} and form a complete set of eigenfunctions for the physical space and parameter regime considered~\cite{Schev}. Both these properties will be essential for obtaining a consistent map between the fields in two different regions. As can be seen from Fig.~\ref{fig1}~{\bf (b)}, in a given region the radiation and SPP DRs never cross; a manifestation of the orthogonality of the underlying wavefunctions. Thus mode-matching conditions cannot be met. As a result, it is not possible for SPP and radiation excitations to directly couple to each other in the same region. However, the situation changes with the introduction of an interface where two regions with different physical properties are joined together. Depending on the material properties, field-matching and energy conservation, coupling between SPPs and radiation within the same and across two different regions can occur. Before we treat such a situation in detail, we first ensure that all the excitations involved (as defined by Eqs.~(\ref{vecr})~and~(\ref{vecp})) are normalized correctly.  

\section{Normalization} 

A standard approach in classical coupled mode theory is to use the electromagnetic reciprocity and Poynting's theorems to ensure that the various fields are normalized correctly with respect to energy transfer in the direction normal to an interface~\cite{SnyderLove}. In the quantum case we check the normalization by starting from the quantized version of the Poynting vector~\cite{Loudon}, which remains valid for the parameter regime considered here, corresponding to a weakly dispersive media~\cite{Land,Huttner,Loudon2,Zayats,Berini}. The quantized Poynting vector is given by
\bea
\hat{\bf S}_{\mu i}({\bf r},t)&=& \hat{\bf E}_{\mu i}^-({\bf r},t)\times \hat{\bf H}_{\mu i}^+({\bf r},t)  -   \hat{\bf H}_{\mu i}^-({\bf r},t) \times \hat{\bf E}_{\mu i}^+({\bf r},t). \label{Poynq}
\eea 
Here, $\mu \in \{r,p\}$ corresponds to radiation ($r\in\{{\textit {\footnotesize TM}},{\textit {\footnotesize TE}}\}$) or SPP ($p$) excitations.
The quantized fields $\hat{\bf E}_{\mu i}({\bf
r},t)$ and $\hat{\bf H}_{\mu i}({\bf
r},t)$ can be obtained in the usual way from the vector potential $\hat{\bf A}_{\mu i}({\bf
r},t)$ defined in Eqs.~(\ref{vecr})~and~(\ref{vecp}) with the relations $\hat{\bf E}_{\mu i}({\bf
r},t)=-\partial_t\hat{\bf A}_{\mu i}({\bf
r},t)$ and $\hat{\bf H}_{\mu i}({\bf
r},t)=\mu_0^{-1}(\nabla \times \hat{\bf A}_{\mu i}({\bf
r},t))$. The same can be carried out for the quantized fields in region $j$.
The $\pm$ superscript corresponds to the positive and negative frequency parts of a given field operator, the explicit form of which are provided in Appendix A. 
Upon substitution of the relevant fields into Eq.~(\ref{Poynq}) and taking the $\hat{\bf x}$ component (direction normal to the interface) integrated over time and $y$-$z$ cross-section at the point $x=0$, one finds
\bea
& &\hskip-0.5cm \iiint  \hat{\bf S}_{\mu i}( {\bf r},t)  \cdot \hat{\bf x}\,{\rm d}t{\rm d}y{\rm d}z \nonumber \\
&&\quad =\frac{1}{(2\pi)^2}  \iint {\rm d}{\bf k}_i {\rm d}{\bf k}'_i \int_0^{q_{\rm cut}}\int_0^{q_{\rm cut}} {\rm d}q_i {\rm d}q'_i   \iiint {\rm d}t{\rm d}y{\rm d}z \,   \nonumber \\ 
&& \quad\quad \left( \hat{\bf E}_{\mu i}^-({{{\bf k}_i},q_i},{\bf r},t)\times \hat{\bf H}_{\mu i}^+({{{\bf k}'_i},q'_i},{\bf r},t)  \right.  \nonumber \\
&&\quad\quad \quad \quad \quad  \left.  -    \hat{\bf H}_{\mu i}^-({{{\bf k}_i},q_i},{\bf r},t)\times \hat{\bf E}_{\mu i}^+({{{\bf k}'_i},q'_i},{\bf r},t)     \right) \cdot  \hat{\bf x}, \label{poynnorm}
\eea 
where for the SPP excitations ($\mu=p$) the parameter $q_i$ and its integration are removed from the field definitions \cite{ft1}. The limits on all integrals are $-\infty \to +\infty$ unless stated otherwise.

The reciprocity theorem of electrodynamics imposes an orthogonality relation for the wavefunctions of the electromagnetic excitations with the same time dependent harmonic evolution \cite{SnyderLove}. If we
consider that the radiation and SPP excitations are not damped in their direction of propagation, i.e. $k_{x,i}$ and $k_{y,i}$ are both real (we will return to this point in more detail in Section VI), the orthogonality condition is given by
\bea
&& \hskip-1cm\iiint  {\bf E}_{\mu i}^-({{{\bf k}_i},q_i},{\bf r},t)\times {\bf H}_{\nu i}^+({{{\bf k}_i'},q_i'},{\bf r},t)  \cdot \hat{\bf x}\,{\rm d}t{\rm d}y{\rm d}z \nonumber \\
&& \quad \quad \quad = |{\cal N}_{\mu i}({\bf k}_i,q_i)|^2\delta{({\bf k}_i - {\bf k}_i')}\delta{(q_i - q_i')}\delta_{\mu \nu},
\label{Recipq}
\eea
where $|{\cal N}_{\mu i}({\bf k}_i,q_i)|^2=2\pi^2 \hbar \omega_i$ is found using the relations provided in Appendix A~\cite{omegatrans}.
This expression allows us to recast Eq.~(\ref{poynnorm}) for forward propagating excitations ($k_{x,i}$: $+{\rm ve}$) as
\be
\iiint  \hat{\bf S}_{r i}( {\bf r},t)  \cdot \hat{\bf x}\,{\rm d}t{\rm d}y{\rm d}z =\iint_{0}^{q_{\rm cut}} {\rm d}{\bf k}_i {\rm d}q_i \,  \hbar \omega_i \, \hat{n}_r({\bf k}_i,q_i) ,
\label{energyrad}
\ee
for the radiation and
\be
\iiint  \hat{\bf S}_{p i}( {\bf r},t)  \cdot \hat{\bf x}\,{\rm d}t{\rm d}y{\rm d}z = \int {\rm d}{\bf k}_i \,  \hbar \omega_i \, \hat{n}_p({\bf k}_i) ,
\label{energyspp}
\ee
for the SPPs. Here $\hat{n}_p=\hat{a}^\dag\hat{a}$ and $\hat{n}_r=\hat{A}_r^\dag\hat{A}_r$ are the particle number operators. Physically, the right hand sides of Eqs.~(\ref{energyrad})~and~(\ref{energyspp}) represent the sum of the excitation energies of the forward propagating field modes above their zero-point value~\cite{Loudon2}. Thus they represent the total radiation and SPP energy that flows through the $y$-$z$ plane in the $+\hat{\bf x}$ direction at $x=0$. Similar equations hold for the backward propagating excitations. With the fields normalized correctly with respect to energy flow across the interface, we can now proceed to develop the formalism for the coupling between the excitations.

\section{Quantized field matching}
We start by using the $\hat{\bf E}^\pm$ and $\hat{\bf H}^\pm$ fields to derive the coupling between SPP and radiation excitations on either side of the interface shown in Fig.~\ref{fig1}~{\bf (a)}. In doing this, we extend a recent classical study on coupled hybrid modes~\cite{Oult} to the quantum domain. At the same time, the theory we develop here is more general than Ref.~\cite{Oult} in the sense that it includes SPP scattering at an angle of incidence to the interface. This is not a straightforward extension, even in the classical case and requires a great deal of care to be taken in the derivation due to inter-polarization coupling. 

For simplicity, the wavefunctions of all excitations are taken as plane waves, extending infinitely in the direction perpendicular to that of propagation. A beam-width can be imposed in a straightforward manner in order to make them finite and closer in description to an experiment, however this does not result in any change to the underlying theory~\cite{Loudon,Blow,TSPP,DSPP}.

The total normalized quantized electric field on side $i$ consisting of SPPs and radiation can be written as
\bea
&&\hskip-0.5cm \hat{\bf E}_i({\bf r}, t)=\int\limits_{\stackrel{k_{y,i} +{\rm ve}}{ {\scriptscriptstyle k_{x,i}+{\rm ve}}}}[\hat{\bf E}_{pi}^{+f}({\bf r},{\bf k}_i, t)+\hat{\bf E}_{pi}^{-f}({\bf r},{\bf k}_i, t)]\,{\rm d}{\bf k}_i \nonumber \\
&&+\int\limits_{\stackrel{k_{y,i} +{\rm ve}}{ {\scriptscriptstyle k_{x,i} +{\rm ve}}}}[\hat{\bf E}_{pi}^{+b}({\bf r},{\bf k}_i, t)+\hat{\bf E}_{pi}^{-b}({\bf r},{\bf k}_i, t)]\,{\rm d}{\bf k}_i \nonumber \\
&&+\int\limits_{\stackrel{k_{y,i} -{\rm ve}}{ {\scriptscriptstyle k_{x,i} \pm{\rm ve}}}}[\hat{\bf E}_{pi}^{+o}({\bf r},{\bf k}_i, t)+\hat{\bf E}_{pi}^{-o}({\bf r},{\bf k}_i, t)]\,{\rm d}{\bf k}_i\nonumber \\
&&\hskip-0.0cm+\sum_{r}\bigg[\iint\limits_{\stackrel{k_{y,i} +{\rm ve}}{ {\scriptscriptstyle k_{x,i} +{\rm ve}}}}[\hat{\bf E}_{ri}^{+f}({\bf r},{\bf k}_i, q_i, t)+\hat{\bf E}_{ri}^{-f}({\bf r},{\bf k}_i, q_i,  t)]\,{\rm d}{\bf k}_i{\rm d}q_i \nonumber \\
&&\hskip-0.0cm+\iint\limits_{\stackrel{k_{y,i} +{\rm ve}}{ {\scriptscriptstyle k_{x,i} +{\rm ve}}}}[\hat{\bf E}_{ri}^{+b}({\bf r},{\bf k}_i, q_i, t)+\hat{\bf E}_{ri}^{-b}({\bf r},{\bf k}_i, q_i,  t)]\,{\rm d}{\bf k}_i{\rm d}q_i\nonumber \\
&&\hskip-0.0cm+\iint\limits_{\stackrel{k_{y,i} -{\rm ve}}{ {\scriptscriptstyle k_{x,i} \pm{\rm ve}}}}[\hat{\bf E}_{ri}^{+o}({\bf r},{\bf k}_i, q_i, t)+\hat{\bf E}_{ri}^{-o}({\bf r},{\bf k}_i, q_i,  t)]\,{\rm d}{\bf k}_i{\rm d}q_i \bigg],
\eea
where the ${\bf k}_{i}$ wavevector integral has been split into three parts for the SPPs and the radiation (TM and TE). The range of the components of the wavevectors for each of these parts is specified below the integral sign. Here and elsewhere the superscripts $f$ and $b$ on the operators denote `forward' and `backward' propagating excitations, whereas $o$ denotes excitations with all `other' propagation directions. As a special case, the superscript $b$ denotes a reversing of the $k_{x}$ component in an operator and its associated wavefunction. This allows the integrals for the forward and backward excitations to be combined later. No change to operators or wavefunctions for the $f$ and $o$ excitations should be made. 
For the quantized magnetic field we have
\bea
&&\hskip-0.5cm\hat{\bf H}_i({\bf r}, t)= \int\limits_{\stackrel{k_{y,i} +{\rm ve}}{ {\scriptscriptstyle k_{x,i} +{\rm ve}}}}[\hat{\bf H}_{pi}^{+f}({\bf r},{\bf k}_i, t)+\hat{\bf H}_{pi}^{-f}({\bf r},{\bf k}_i, t)]\,{\rm d}{\bf k}_i\nonumber \\
&&+\int\limits_{\stackrel{k_{y,i} +{\rm ve}}{ {\scriptscriptstyle k_{x,i} +{\rm ve}}}}[\hat{\bf H}_{pi}^{+b}({\bf r},{\bf k}_i, t)+\hat{\bf H}_{pi}^{-b}({\bf r},{\bf k}_i, t)]\,{\rm d}{\bf k}_i \nonumber \\
&&+\int\limits_{\stackrel{k_{y,i} -{\rm ve}}{ {\scriptscriptstyle k_{x,i} \pm{\rm ve}}}}[\hat{\bf H}_{pi}^{+o}({\bf r},{\bf k}_i, t)+\hat{\bf H}_{pi}^{-o}({\bf r},{\bf k}_i, t)]\,{\rm d}{\bf k}_i\nonumber \\
&&\hskip-0.0cm+\sum_{r}\bigg[\iint\limits_{\stackrel{k_{y,i} +{\rm ve}}{ {\scriptscriptstyle k_{x,i} +{\rm ve}}}}[\hat{\bf H}_{ri}^{+f}({\bf r},{\bf k}_i, q_i, t)+\hat{\bf H}_{ri}^{-f}({\bf r},{\bf k}_i, q_i,  t)]\,{\rm d}{\bf k}_i{\rm d}q_i \nonumber \\
&&\hskip-0.0cm+\iint\limits_{\stackrel{k_{y,i} +{\rm ve}}{ {\scriptscriptstyle k_{x,i} +{\rm ve}}}}[\hat{\bf H}_{ri}^{+b}({\bf r},{\bf k}_i, q_i, t)+\hat{\bf H}_{ri}^{-b}({\bf r},{\bf k}_i, q_i,  t)]\,{\rm d}{\bf k}_i{\rm d}q_i\nonumber \\
&&\hskip-0.0cm+\iint\limits_{\stackrel{k_{y,i} -{\rm ve}}{ {\scriptscriptstyle k_{x,i} \pm{\rm ve}}}}[\hat{\bf H}_{ri}^{+o}({\bf r},{\bf k}_i, q_i, t)+\hat{\bf H}_{ri}^{-o}({\bf r},{\bf k}_i, q_i,  t)]\,{\rm d}{\bf k}_i{\rm d}q_i\bigg]. 
\eea
Similar expressions can be written for the quantized fields on side $j$. We now match the transverse part of the fields on either side at point $x=0$ as follows
\bea
\hat{\mathscr E}_i({\bf r},t)&=&\hat{\mathscr E}_j({\bf r},t), \label{transa}\\
\hat{\mathscr H}_i({\bf r},t)&=&\hat{\mathscr H}_j({\bf r},t),
\label{transb}
\eea 
where $\hat{\mathscr E}_i({\bf r},t)=(\hat{\bf E}_i({\bf r},t)\cdot\hat{\bf y})\, \hat{\bf y}+(\hat{\bf E}_i({\bf r},t)\cdot\hat{\bf z})\, \hat{\bf z}$ and $\hat{\mathscr H}_i({\bf r},t)=(\hat{\bf H}_i({\bf r},t)\cdot\hat{\bf y})\, \hat{\bf y}+(\hat{\bf H}_i({\bf r},t)\cdot\hat{\bf z})\, \hat{\bf z}$~\cite{QMequiv}. By noting the following relations for the transverse component of the wavefunctions: ${\mathscr E}_{pi}^{+b}({\bf r},{\bf k}_i,t)={\mathscr E}_{pi}^{+f}({\bf r},{\bf k}_i,t)$, ${\mathscr E}_{{\textit {\tiny TM}}\,i}^{+b}({\bf r},{\bf k}_i,q_i,t)={\mathscr E}_{{\textit {\tiny TM}}\,i}^{+f}({\bf r},{\bf k}_i,q_i,t)$, ${\mathscr E}_{{\textit {\tiny TE}}\,i}^{+b}({\bf r},{\bf k}_i,q_i,t)=-{\mathscr E}_{{\textit {\tiny TE}}\,i}^{+f}({\bf r},{\bf k}_i,q_i,t)$, ${\mathscr H}_{pi}^{+b}({\bf r},{\bf k}_i,t)=-{\mathscr H}_{pi}^{+f}({\bf r},{\bf k}_i,t)$, ${\mathscr H}_{{\textit {\tiny TM}}\,i}^{+b}({\bf r},{\bf k}_i,q_i,t)=-{\mathscr H}_{{\textit {\tiny TM}}\,i}^{+f}({\bf r},{\bf k}_i,q_i,t)$ and ${\mathscr H}_{{\textit {\tiny TE}}\,i}^{+b}({\bf r},{\bf k}_i,q_i,t)={\mathscr H}_{{\textit {\tiny TE}}\,i}^{+f}({\bf r},{\bf k}_i,q_i,t)$ we form a coupled equation for the quantized electric field
\bea
& &\hskip-0.2cm \int\limits_{\stackrel{k_{y,i} +{\rm ve}}{ {\scriptscriptstyle k_{x,i} +{\rm ve}}}}[\hat{a}^f({\bf k}_i)+\hat{a}^b({\bf k}_i)]{\mathscr E}_{pi}^{+f}({\bf r},{\bf k}_i,t)\,{\rm d}{\bf k}_i\nonumber \\
&&\hskip-0.2cm+\int\limits_{\stackrel{k_{y,i} -{\rm ve}}{ {\scriptscriptstyle k_{x,i} \pm{\rm ve}}}}[\hat{a}^o({\bf k}_i)]{\mathscr E}_{pi}^{+o}({\bf r},{\bf k}_i,t)\,{\rm d}{\bf k}_i \nonumber \\
&&\hskip-0.2cm+\sum_{r}\bigg[\iint\limits_{\stackrel{k_{y,i} +{\rm ve}}{ {\scriptscriptstyle k_{x,i} +{\rm ve}}}}[\hat{A}_r^f({\bf k}_i,q_i )\pm  \hat{A}_r^b({\bf k}_i,q_i)]{\mathscr E}_{ri}^{+f}({\bf r},{\bf k}_i,q_i,t)\,{\rm d}{\bf k}_i{\rm d}q_i\nonumber \\
&&\hskip-0.2cm+\iint\limits_{\stackrel{k_{y,i} -{\rm ve}}{ {\scriptscriptstyle k_{x,i} \pm{\rm ve}}}}[\hat{A}_r^o({\bf k}_i,q_i)]{\mathscr E}_{ri}^{+o}({\bf r},{\bf k}_i,q_i,t)\,{\rm d}{\bf k}_i{\rm d}q_i \bigg]+ H.c. \nonumber \\
&&\hskip-0.2cm= \int\limits_{\stackrel{k_{y,j} +{\rm ve}}{ {\scriptscriptstyle k_{x,j} +{\rm ve}}}}[\hat{b}^f({\bf k}_j)+\hat{b}^b({\bf k}_j)]{\mathscr E}_{pj}^{+f}({\bf r},{\bf k}_j,t)\,{\rm d}{\bf k}_j\nonumber \\
&&\hskip-0.2cm+\int\limits_{\stackrel{k_{y,j} -{\rm ve}}{ {\scriptscriptstyle k_{x,j} \pm{\rm ve}}}}[\hat{b}^o({\bf k}_j)]{\mathscr E}_{pj}^{+o}({\bf r},{\bf k}_j,t)\,{\rm d}{\bf k}_j \nonumber \\
&&\hskip-0.2cm+\sum_{r}\bigg[\iint\limits_{\stackrel{k_{y,j} +{\rm ve}}{ {\scriptscriptstyle k_{x,j} +{\rm ve}}}}[\hat{B}_r^f({\bf k}_j,q_j) \pm  \hat{B}_r^b({\bf k}_j,q_j)]{\mathscr E}_{rj}^{+f}({\bf r},{\bf k}_j,q_j,t)\,{\rm d}{\bf k}_j{\rm d}q_j\nonumber \\
&&\hskip-0.2cm+\iint\limits_{\stackrel{k_{y,j} -{\rm ve}}{ {\scriptscriptstyle k_{x,j} \pm{\rm ve}}}}[\hat{B}_r^o({\bf k}_j,q_j)]{\mathscr E}_{rj}^{+o}({\bf r},{\bf k}_j,q_j,t)\,{\rm d}{\bf k}_j{\rm d}q_j\bigg] + H.c.
\label{electrict}
\eea
and another for the quantized magnetic field
\bea
& &\hskip-0.3cm \int\limits_{\stackrel{k_{y,i} +{\rm ve}}{ {\scriptscriptstyle k_{x,i} +{\rm ve}}}}[\hat{a}^f({\bf k}_i)-\hat{a}^b({\bf k}_i)]{\mathscr H}_{pi}^{+f}({\bf r},{\bf k}_i,t)\,{\rm d}{\bf k}_i\nonumber \\
&&\hskip-0.3cm+\int\limits_{\stackrel{k_{y,i} -{\rm ve}}{ {\scriptscriptstyle k_{x,i} \pm{\rm ve}}}}[\hat{a}^o({\bf k}_i)]{\mathscr H}_{pi}^{+o}({\bf r},{\bf k}_i,t)\,{\rm d}{\bf k}_i \nonumber \\
&&\hskip-0.3cm+\sum_{r}\bigg[\iint\limits_{\stackrel{k_{y,i} +{\rm ve}}{ {\scriptscriptstyle k_{x,i} +{\rm ve}}}}[\hat{A}_r^f({\bf k}_i,q_i)   \mp   \hat{A}_r^b({\bf k}_i,q_i)]{\mathscr H}_{ri}^{+f}({\bf r},{\bf k}_i,q_i,t)\,{\rm d}{\bf k}_i{\rm d}q_i\nonumber \\
&&\hskip-0.3cm+\iint\limits_{\stackrel{k_{y,i} -{\rm ve}}{ {\scriptscriptstyle k_{x,i} \pm{\rm ve}}}}[\hat{A}_r^o({\bf k}_i,q_i)]{\mathscr H}_{ri}^{+o}({\bf r},{\bf k}_i,q_i,t)\,{\rm d}{\bf k}_i{\rm d}q_i \bigg] + H.c. \nonumber \\
&&\hskip-0.3cm= \int\limits_{\stackrel{k_{y,j} +{\rm ve}}{ {\scriptscriptstyle k_{x,j} +{\rm ve}}}}[\hat{b}^f({\bf k}_j)-\hat{b}^b({\bf k}_j)]{\mathscr H}_{pj}^{+f}({\bf r},{\bf k}_j,t)\,{\rm d}{\bf k}_j\nonumber \\
&&\hskip-0.3cm+\int\limits_{\stackrel{k_{y,j} -{\rm ve}}{ {\scriptscriptstyle k_{x,j} \pm{\rm ve}}}}[\hat{b}^o({\bf k}_j)]{\mathscr H}_{pj}^{+o}({\bf r},{\bf k}_j,t)\,{\rm d}{\bf k}_j \nonumber \\
&&\hskip-0.3cm+\sum_{r}\bigg[\iint\limits_{\stackrel{k_{y,j} +{\rm ve}}{ {\scriptscriptstyle k_{x,j} +{\rm ve}}}}[\hat{B}_r^f({\bf k}_j,q_j)   \mp   \hat{B}_r^b({\bf k}_j,q_j)]{\mathscr H}_{rj}^{+f}({\bf r},{\bf k}_j,q_j,t)\,{\rm d}{\bf k}_j{\rm d}q_j\nonumber \\
&&\hskip-0.3cm+\iint\limits_{\stackrel{k_{y,j} -{\rm ve}}{ {\scriptscriptstyle k_{x,j} \pm{\rm ve}}}}[\hat{B}_r^o({\bf k}_j,q_j)]{\mathscr H}_{rj}^{+o}({\bf r},{\bf k}_j,q_j,t)\,{\rm d}{\bf k}_j{\rm d}q_j\bigg]     + H.c.  \, \,\,
\label{magnetict}
\eea
Here the top (bottom) sign of the symbols $\pm$ and $\mp$ corresponds to TM (TE) radiation. These two equations form the basis on which the coupling between the different excitations can be calculated. However, first they must be broken down into a more convenient form. Taking Eq.~(\ref{electrict}) and post-multiplying both sides by $\times {\mathscr H}_{pi}^{-f}({\bf r},{\bf k}_i',t)$, then integrating over $\iiint \cdot\hat{\bf x}\,{\rm d}t{\rm d}y{\rm d}z$ (using the orthogonality relation given in Eq.~(\ref{Recipq}) to drop terms associated with orthogonal wavefunctions) and finally relabeling ${\bf k}_i' \to {\bf k}_i$, selects out the coupled annihilation operator equation
\bea
&&\hskip-0.6cm [\hat{a}^f({\bm \omega}_i)+\hat{a}^b({\bm \omega}_i)] {\cal C}^{ii}_{pp}({\bf k}_i;{\bf k}_i) =[\hat{b}^f({\bm \omega}_j)+\hat{b}^b({\bm \omega}_j)]{\cal C}^{ji}_{pp}({\bf k}_j;{\bf k}_i)\nonumber \\
&&\hskip-0.5cm +\sum_{r} \int[\hat{B}_r^f({\bm \omega}_j,q_j)   \pm   \hat{B}_r^b({\bm \omega}_j,q_j)]{\cal C}^{ji}_{rp}({\bf k}_j,q_j;{\bf k}_i){\rm d}q_j,
\label{couple1}
\eea
where the coupling coefficients are given by
\bea
&&\hskip-0.5cm{\cal C}^{ji}_{\mu\nu}({\bf k}_j,q_j;{\bf k}_i,q_i)= \label{couplcoeff}\\
&&\hskip-0.1cm {\cal N}\iiint[{\mathscr E}_{\mu j}^{+f}({\bf r},{\bf k}_j,q_j,t)\times {\mathscr H}_{\nu i}^{-f}({\bf r},{\bf k}_i,q_i,t)]\cdot \hat{\bf x}\,{\rm d}t{\rm d}y{\rm d}z \nonumber \\
&&\hskip-0.1cm\equiv {\cal N}\iiint[{\bf E}_{\mu j}^{+f}({\bf r},{\bf k}_j,q_j,t)\times {\bf H}_{\nu i}^{-f}({\bf r},{\bf k}_i,q_i,t)]\cdot \hat{\bf x}\,{\rm d}t{\rm d}y{\rm d}z, \nonumber
\eea
with ${\cal N}=(2 \pi^2 \hbar \omega_i)^{-1}$. Note that we have introduced a vector ${\bm \omega}_i=(\omega_i,k_{y,i})$ in Eq.~(\ref{couple1}). This is due to the $k_{x,i}$ component of the wavevector ${\bf k}_i$ in the operators being transformed into the frequency domain during the evaluation of the coupling coefficients. For SPPs in a given region, once $k_y$ is set, $\omega$ determines $k_x$ via the dispersion relation. For radiation the same is true, once $k_y$ and $q$ are set, $\omega$ determines $k_x$ via the corresponding dispersion relation. Therefore $k_x$ is no longer a free parameter for the operators. The full analytical form of the couplings and further details on the domain transfer are provided in Appendix B. 

Next, we take Eq.~(\ref{electrict}) and post-multiply both sides by $\times {\mathscr H}_{{\textit {\tiny TM}}\,i}^{-f}({\bf r},{\bf k}_i',q_i',t)$, integrating over $\iiint \cdot\hat{\bf x}\,{\rm d}t{\rm d}y{\rm d}z$ and finally relabeling ${\bf k}_i' \to {\bf k}_i$ and $q_i' \to q_i$. This selects out the coupled operator equation
\bea
&&\hskip-0.8cm  [\hat{A}_{{\textit {\tiny TM}}}^f({\bm \omega}_i,q_i)+\hat{A}_{{\textit {\tiny TM}}}^b({\bm \omega}_i,q_i)]  {\cal C}^{ii}_{{\textit {\tiny TM}}\,{\textit {\tiny TM}}}({\bf k}_i,q_i;{\bf k}_i,q_i)  \nonumber \\ 
&&\hskip-0.5cm =[\hat{b}^f({\bm \omega}_j)+\hat{b}^b({\bm \omega}_j)]{\cal C}^{ji}_{p\,{{\textit {\tiny TM}}}}({\bf k}_j;{\bf k}_i,q_i)\nonumber \\
&&\hskip-0.5cm+  \sum_{r}  \int[\hat{B}_r^f({\bm \omega}_j,q_j)   \pm  \hat{B}_r^b({\bm \omega}_j,q_j)]{\cal C}^{ji}_{r\,{\textit {\tiny TM}}}({\bf k}_j,q_j;{\bf k}_i,q_i){\rm d}q_j.
\label{couple2TM}
\eea
Post-multiplying both sides by $\times {\mathscr H}_{{\textit {\tiny TE}}\,i}^{-f}({\bf r},{\bf k}_i',q_i',t)$ instead leads to
\bea
&&\hskip-0.8cm [\hat{A}_{{\textit {\tiny TE}}}^f({\bm \omega}_i,q_i)   -   \hat{A}_{{\textit {\tiny TE}}}^b({\bm \omega}_i,q_i)  ]  {\cal C}^{ii}_{{\textit {\tiny TE}}\,{\textit {\tiny TE}}}({\bf k}_i,q_i;{\bf k}_i,q_i)  \nonumber \\ 
&&\hskip-0.5cm =[\hat{b}^f({\bm \omega}_j)+\hat{b}^b({\bm \omega}_j)]{\cal C}^{ji}_{p\,{{\textit {\tiny TE}}}}({\bf k}_j;{\bf k}_i,q_i)\nonumber \\
&&\hskip-0.5cm+  \sum_{r}  \int[\hat{B}_r^f({\bm \omega}_j,q_j)   \pm  \hat{B}_r^b({\bm \omega}_j,q_j)]{\cal C}^{ji}_{r \, {\textit {\tiny TE}}}({\bf k}_j,q_j;{\bf k}_i,q_i){\rm d}q_j.
\label{couple2TE}
\eea
Then, we take Eq.~(\ref{magnetict}) and pre-multiply both sides by ${\mathscr E}_{pj}^{-f}({\bf r},{\bf k}_j',t)\times$, integrate over $\iiint \cdot\hat{\bf x}\,{\rm d}t{\rm d}y{\rm d}z$ and relabel ${\bf k}_j' \to {\bf k}_j$. This selects out the coupled operator equation
\bea
&&\hskip-0.6cm [ \hat{b}^f({\bm \omega}_j)-\hat{b}^b({\bm \omega}_j) ]{{\cal C}^{jj}_{pp}}^{*}({\bf k}_j;{\bf k}_j) =[\hat{a}^f({\bm \omega}_i)-\hat{a}^b({\bm \omega}_i)]{{\cal C}^{ji}_{pp}}^{*}({\bf k}_j;{\bf k}_i)\nonumber \\
&&\hskip-0.3cm+\sum_{r} \int[\hat{A}_r^f({\bm \omega}_i,q_i)    \mp     \hat{A}_r^b({\bm \omega}_i,q_i)]{{\cal C}^{ji}_{pr}}^{*}({\bf k}_j;{\bf k}_i,q_i){\rm d}q_i.
\label{couple3}
\eea
Next, we take Eq.~(\ref{magnetict}) and pre-multiply both sides by ${\mathscr E}_{{\textit {\tiny TM}}\,j}^{-f}({\bf r},{\bf k}_j',q_j',t)\times$, integrate over $\iiint \cdot\hat{\bf x}\,{\rm d}t{\rm d}y{\rm d}z$ and relabel ${\bf k}_j' \to {\bf k}_j$ and $q_j' \to q_j$. This selects out the coupled operator equation
\bea
&&\hskip-0.7cm [\hat{B}_{{\textit {\tiny TM}}}^f({\bm \omega}_j,q_j)-\hat{B}_{{\textit {\tiny TM}}}^b({\bm \omega}_j,q_j) ]{{\cal C}^{jj\,*}_{{{\textit {\tiny TM}}}\,{{\textit {\tiny TM}}}} }({\bf k}_j,q_j;{\bf k}_j,q_j) \nonumber \\
&&\hskip-0.3cm =[\hat{a}^f({\bm \omega}_i)-\hat{a}^b({\bm \omega}_i)] {{\cal C}^{ji\,*}_{{{\textit {\tiny TM}}}\,p} } ({\bf k}_j,q_j;{\bf k}_i) \nonumber \\
&&\hskip-0.3cm+ \sum_{r}  \int[\hat{A}_r^f({\bm \omega}_i,q_i)    \mp   \hat{A}_r^b({\bm \omega}_i,q_i)]{{\cal C}^{ji\,*}_{{{\textit {\tiny TM}}}\,r} }({\bf k}_j,q_j;{\bf k}_i,q_i){\rm d}q_i.
\label{couple4TM}
\eea
Finally, pre-multiplying both sides by ${\mathscr E}_{{\textit {\tiny TE}}\,i}^{-f}({\bf r},{\bf k}_j',q_j',t)\times$ instead leads to
\bea
&&\hskip-0.7cm [\hat{B}_{{\textit {\tiny TE}}}^f({\bm \omega}_j,q_j) + \hat{B}_{{\textit {\tiny TE}}}^b({\bm \omega}_j,q_j) ]{{\cal C}^{jj\,*}_{{{\textit {\tiny TE}}}\,{{\textit {\tiny TE}}}} }({\bf k}_j,q_j;{\bf k}_j,q_j) \nonumber \\
&&\hskip-0.3cm=[\hat{a}^f({\bm \omega}_i) - \hat{a}^b({\bm \omega}_i)] {{\cal C}^{ji\,*}_{{{\textit {\tiny TE}}}\,p} } ({\bf k}_j,q_j;{\bf k}_i) \nonumber \\
&&\hskip-0.3cm+ \sum_{r}  \int[\hat{A}_r^f({\bm \omega}_i,q_i)    \mp   \hat{A}_r^b({\bm \omega}_i,q_i)]{{\cal C}^{ji\,*}_{{{\textit {\tiny TE}}}\,r} }({\bf k}_j,q_j;{\bf k}_i,q_i){\rm d}q_i.
\label{couple4TE}
\eea
Eqs.~(\ref{couple1}) and (\ref{couple2TM})-(\ref{couple4TE}) represent six coupled equations that relate all the annihilation operators corresponding to the excitations involved in the scattering process. From these equations we are able to form a transfer matrix, as depicted in Fig.~\ref{fig2}~{\bf (a)}. This matrix will then provide a complete description of the quantum dynamics for the scattering process that occurs at the interface. Note that on the right hand side of each of the six coupled equations we have kept coupling coefficients which, by definition of Eqs.~(\ref{Recipq})~and~(\ref{couplcoeff}), should strictly be equal to unity. We keep these coefficients in order to allow for the inclusion of evanescent excitations in our theory which have a normalization equivalent to Eq.~(\ref{Recipq}) up to an overall factor $\pm i$~\cite{Marcuv}, which we will specify when required.

\section{Transfer matrix}
\begin{figure}[t]
\centerline{\psfig{figure=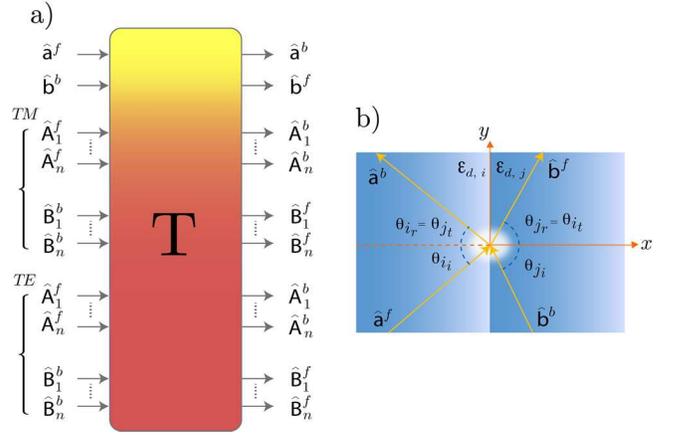,width=8.5cm}}
\caption{(Color online) Transfer matrix and SPP beamsplitter angles. {\bf (a)}: Input-output transfer matrix $T$ linking all the excitations at the interface. {\bf (b)}: Angles involved in the scattering process: the incident angle $\theta_{i_i}$ ($\theta_{j_i}$) of the SPP excitation $\hat{a}^f$ ($\hat{b}^b$) in region $i$ ($j$), the transmitted angle $\theta_{i_t}$ ($\theta_{j_t}$) of the SPP excitation $\hat{b}^f$ ($\hat{a}^b$) in region $j$ ($i$). The relations $\theta_{i_r}=\theta_{j_t}$ and $\theta_{j_r}=\theta_{i_t}$ must hold for the excited outputs to be indistinguishable. This can always be achieved, see text for details.}
\label{fig2}
\end{figure}

In order to form the transfer matrix we take Eqs.~(\ref{couple1}) and (\ref{couple2TM})-(\ref{couple4TE}) and use truncated summations to approximate the $q_i$ and $q_j$ integrals for the radiation excitations~\cite{Oult}. For accuracy and convergence Gaussian quadrature summation~\cite{SA} is employed, where an arbitrary integral can be written as
\be
\hskip-0.005cm \int_{a}^b f(q)\,{\rm d}q=\frac{b-a}{2}\int_{-1}^{1} f(u)\,{\rm d}u=\frac{b-a}{2} \sum_{m=1}^{N+1}w_m f(u_m)\label{Gauss}
\ee
with $u=2q/(b-a)+(a+b)/(a-b)$, the weights $w_m=2/[(1-u_m^2)(P'_{N+1}(u_m))^2]$ and the $u_m$ abscissa chosen to be the roots (zeros) of the $P_{N+1}(u)$ the Legendre polynomial. In our case we have the integral limits $a=0$ and $b=q_{\rm cut}$. Here, care must be taken with the weights. We must ensure that the sum of the modulus squared of a given set of amplitudes, for example $\alpha^b_r({\bm \omega}_i,u_m)$ ($\equiv \alpha^b_{r,m}({\bm \omega}_i,q_i)$), calculated from the coupled equations, for the discretized radiation operators $\hat{A}^b_r({\bm \omega}_i,u_m)$ ($\equiv \hat{A}^b_{r,m}({\bm \omega}_i,q_i)$) leads approximately to the total back-scattered energy, {\it i.e.} $\int |\alpha^b_r({\bm \omega}_i,q_i)|^2\,{\rm d}q_i\simeq\sum_{m=1}^{N+1}\,|\alpha^b_{r,m}({\bm \omega}_i,q_i)|^2$. But, following Eq.~(\ref{Gauss}), we have the relation $\int |\alpha^b_r({\bm \omega}_i,q_i)|^2\,{\rm d}q_i=\sum_{m=1}^{N+1}w_m'\,|\alpha^b_{r,m}({\bm \omega}_i,q_i)|^2$, with $w_m'=w_m\,q_{\rm cut}/2$. Thus we  
rescale all operators involved in integrations, {\it e.g.} $\hat{A}^b_r({\bm \omega}_i,q_i) \to \hat{A}^b_{r,m}({\bm \omega}_i,q_i) \sqrt{w_m'}$ to compensate and write the six coupled operator equations in discretized form as
\bea
&&\hskip-0.45cm [ \hat{a}^f({\bm \omega}_i)+\hat{a}^b({\bm \omega}_i) ]{\cal C}^{ii}_{pp}({\bf k}_i;{\bf k}_i) =[\hat{b}^f({\bm \omega}_j)+\hat{b}^b({\bm \omega}_j)]{\cal C}^{ji}_{pp}({\bf k}_j;{\bf k}_i) \label{coup1} \\
&& \hskip0.1cm+\sum_{r,m=1}^{N+1}[\hat{B}^f_{r,m}({\bm \omega}_j,q_j)  \pm  \hat{B}^b_{r,m}({\bm \omega}_j,q_j)]{\cal C}^{ji}_{rp}({\bf k}_j,u_m;{\bf k}_i)\sqrt{w_m'}, \nonumber \\
&& \hskip-0.45cm [\hat{A}^f_{{\textit {\tiny TM}},n}({\bm \omega}_i,q_i)+\hat{A}^b_{{\textit {\tiny TM}},n}({\bm \omega}_i,q_i) ]{\cal C}^{ii}_{{\textit {\tiny TM}} \, {\textit {\tiny TM}}}({\bf k}_i,u_n;{\bf k}_i,u_n) = \label{coup2}\\
&&\hskip0.45cm[\hat{b}^f({\bm \omega}_j)+\hat{b}^b({\bm \omega}_j)]{\cal C}^{ji}_{p\, {\textit {\tiny TM}}}({\bf k}_j;{\bf k}_i,u_n)\sqrt{w_n'}\nonumber \\
&&\hskip-0.3cm+\sum_{r}  [\hat{B}^f_{r,n}({\bm \omega}_j,q_j) \pm \hat{B}^b_{r,n}({\bm \omega}_j,q_j)]{\cal C}^{ji}_{r \, {\textit {\tiny TM}}}({\bf k}_j,u_n;{\bf k}_i,u_n) \nonumber \\
&&\hskip-0.3cm+\sum_{r,m=1}^{N+1}[\hat{B}^f_{r,m}({\bm \omega}_j,q_j) \pm \hat{B}^b_{r,m}({\bm \omega}_j,q_j)]{\cal C}^{ji}_{r \, {\textit {\tiny TM}}}({\bf k}_j,u_m;{\bf k}_i,u_n)\sqrt{w_m'w_n'}, \nonumber \\
&& \hskip-0.45cm [\hat{A}^f_{{\textit {\tiny TE}},n}({\bm \omega}_i,q_i)-\hat{A}^b_{{\textit {\tiny TE}},n}({\bm \omega}_i,q_i) ]{\cal C}^{ii}_{{\textit {\tiny TE}} \, {\textit {\tiny TE}}}({\bf k}_i,u_n;{\bf k}_i,u_n) = \label{coup3} \\
&&\hskip0.45cm[\hat{b}^f({\bm \omega}_j)+\hat{b}^b({\bm \omega}_j)]{\cal C}^{ji}_{p\, {\textit {\tiny TE}}}({\bf k}_j;{\bf k}_i,u_n)\sqrt{w_n'}\nonumber \\
&&\hskip-0.3cm+\sum_{r}  [\hat{B}^f_{r,n}({\bm \omega}_j,q_j) \pm \hat{B}^b_{r,n}({\bm \omega}_j,q_j)]{\cal C}^{ji}_{r \, {\textit {\tiny TE}}}({\bf k}_j,u_n;{\bf k}_i,u_n) \nonumber \\
&&\hskip-0.3cm+\sum_{r,m=1}^{N+1}[\hat{B}^f_{r,m}({\bm \omega}_j,q_j) \pm \hat{B}^b_{r,m}({\bm \omega}_j,q_j)]{\cal C}^{ji}_{r \, {\textit {\tiny TE}}}({\bf k}_j,u_m;{\bf k}_i,u_n)\sqrt{w_m'w_n'}, \nonumber \\
&&\hskip-0.45cm [\hat{b}^f({\bm \omega}_j)-\hat{b}^b({\bm \omega}_j) ]{{\cal C}^{jj}_{pp}}^{*}({\bf k}_j;{\bf k}_j)=  \label{coup4} \\
&&\hskip0.55cm [\hat{a}^f({\bm \omega}_i)-\hat{a}^b({\bm \omega}_i)]{{\cal C}^{ji}_{pp}}^{*}({\bf k}_j;{\bf k}_i) \nonumber \\
&&\hskip-0.3cm+\sum_{r,m=1}^{N+1}[\hat{A}^f_{r,m}({\bm \omega}_i,q_i) \mp \hat{A}^b_{r,m}({\bm \omega}_i,q_i)]{{\cal C}^{ji}_{pr}}^{*}({\bf k}_j;{\bf k}_i,u_m)\sqrt{w_m'}, \nonumber \\
&& \hskip-0.45cm [\hat{B}^f_{{\textit {\tiny TM}},n}({\bm \omega}_j,q_j)-\hat{B}^b_{{\textit {\tiny TM}},n}({\bm \omega}_j,q_j) ]{{\cal C}^{jj\,*}_{{\textit {\tiny TM}}\,  {\textit {\tiny TM}}}}({\bf k}_j,u_n;{\bf k}_j,u_n) = \label{coup5}\\
&&\hskip0.45cm[\hat{a}^f({\bm \omega}_i)-\hat{a}^b({\bm \omega}_i)]{{\cal C}^{ji\,*}_{{\textit {\tiny TM}}\, p}}({\bf k}_j,u_n;{\bf k}_i)\sqrt{w_n'}\nonumber \\
&&\hskip-0.3cm+\sum_{r}[\hat{A}^f_{r,n}({\bm \omega}_i,q_i)  \mp  \hat{A}^b_{r,n}({\bm \omega}_i,q_i)]{{\cal C}^{ji\,*}_{{\textit {\tiny TM}}\,  r}}({\bf k}_j,u_n;{\bf k}_i,u_n) \nonumber \\
&& \hskip-0.3cm+ \sum_{r,m=1}^{N+1}[\hat{A}^f_{r,m}({\bm \omega}_i,q_i)  \mp  \hat{A}^b_{r,m}({\bm \omega}_i,q_i)]{{\cal C}^{ji\,*}_{{\textit {\tiny TM}}\, r}}({\bf k}_j,u_n;{\bf k}_i,u_m)\sqrt{w_m'w_n'}, \nonumber \\
&& \hskip-0.45cm  [\hat{B}^f_{{\textit {\tiny TE}},n}({\bm \omega}_j,q_j)+\hat{B}^b_{{\textit {\tiny TE}},n}({\bm \omega}_j,q_j)   ]{{\cal C}^{jj\,*}_{{\textit {\tiny TE}}\,  {\textit {\tiny TE}}}}({\bf k}_j,u_n;{\bf k}_j,u_n)= \label{coup6}\\
&&\hskip0.15cm-[\hat{a}^f({\bm \omega}_i)-\hat{a}^b({\bm \omega}_i)]{{\cal C}^{ji\,*}_{{\textit {\tiny TE}}\, p}}({\bf k}_j,u_n;{\bf k}_i)\sqrt{w_n'}\nonumber \\
&&\hskip-0.3cm+\sum_{r}[\hat{A}^f_{r,n}({\bm \omega}_i,q_i)  \mp  \hat{A}^b_{r,n}({\bm \omega}_i,q_i)]{{\cal C}^{ji\,*}_{{\textit {\tiny TE}}\,  r}}({\bf k}_j,u_n;{\bf k}_i,u_n) \nonumber \\
&& \hskip-0.3cm+\sum_{r,m=1}^{N+1}[\hat{A}^f_{r,m}({\bm \omega}_i,q_i)  \mp  \hat{A}^b_{r,m}({\bm \omega}_i,q_i)]{{\cal C}^{ji\,*}_{{\textit {\tiny TE}}\, r}}({\bf k}_j,u_n;{\bf k}_i,u_m)\sqrt{w_m'w_n'}, \nonumber 
\eea
where we have chosen $u_m$ ($u_n$) to represent $q_j$ ($q_i$) on the first three equations and vice-versa for the second three equations in the coupling coefficients. Doing this allows us to write the coupled equations in a more compact form as four matrix equations
\bea
\hat{A}_{{\textit {\tiny TM}},n}^f+\hat{A}_{{\textit {\tiny TM}},n}^b= \sum_{r} \sum_{m=0}^{N+1}(\hat{B}_{r,m}^f  \pm  \hat{B}_{r,m}^b)(D_{r\, {\textit {\tiny TM}}})_{mn} \label{couple5TM}\\
\hat{B}_{{\textit {\tiny TM}},n}^f-\hat{B}_{{\textit {\tiny TM}},n}^b=\sum_{r}\sum_{m=0}^{N+1}(\hat{A}_{r,m}^f   \mp  \hat{A}_{r,m}^b) ({F^\dag_{{\textit {\tiny TM}}\,r}  } )_{mn} \label{couple6TM} \\
\hat{A}_{{\textit {\tiny TE}},n}^f-\hat{A}_{{\textit {\tiny TE}},n}^b= \sum_{r} \sum_{m=0}^{N+1}(\hat{B}_{r,m}^f  \pm  \hat{B}_{r,m}^b)(D_{r\, {\textit {\tiny TE}}})_{mn} \label{couple5TE}\\
\hat{B}_{{\textit {\tiny TE}},n}^f+\hat{B}_{{\textit {\tiny TE}},n}^b=\sum_{r}\sum_{m=0}^{N+1}(\hat{A}_{r,m}^f   \mp  \hat{A}_{r,m}^b)({F^\dag_{{\textit {\tiny TE}}\,r } })_{mn}. \label{couple6TE} 
\eea 
Here, Eqs.~(\ref{coup1}) and  (\ref{coup2}) lead to Eq.~(\ref{couple5TM}),  Eqs.~(\ref{coup4}) and  (\ref{coup5}) lead to Eq.~(\ref{couple6TM}), Eq.~(\ref{coup3}) gives Eq.~(\ref{couple5TE}) and Eq.~(\ref{coup6}) gives Eq.~(\ref{couple6TE}), where
\bea
&&\hat{A}_{{\textit {\tiny TM}},n}^{f,b}:=\{\hat{a}^{f,b}({\bm \omega}_i)|\,n=0\}, \label{SPPfromTM} \\ &&\hat{A}_{{\textit {\tiny TE}},n}^{f,b}:=\{0|\,n=0\},  \nonumber \\&& \hat{A}_{r,n}^{f,b}:=\{\hat{A}_{r,n}^{f,b}({\bm \omega}_i,q_i)\equiv \hat{A}_r^{f,b}({\bm \omega}_i,u_{n})|\,n>0\}, \nonumber \\ &&\hat{B}_{{\textit {\tiny TM}},n}^{f,b}:=\{\hat{b}^{f,b}({\bm \omega}_j)|\,n=0\},  \nonumber  \\ &&\hat{B}_{{\textit {\tiny TE}},n}^{f,b}:=\{0|\,n=0\},  \nonumber \\&& \hat{B}_{r,n}^{f,b}:=\{\hat{B}_{r,n}^{f,b}({\bm \omega}_j,q_j)\equiv \hat{B}_r^{f,b}({\bm \omega}_j,u_{n})|\,n>0\}   , \nonumber \eea
with the $D$ and $F$ matrices given in Appendix C. From Eqs.~(\ref{couple5TM})-(\ref{couple6TE}) we can finally write the transfer matrix, shown in Fig.~\ref{fig2}~{\bf (a)}, for the operators at the interface as
\be 
\left(
\begin{array}{c}
\hat{A}_{{\textit {\tiny TM}}}^{b} \\
\hat{B}_{{\textit {\tiny TM}}}^{f} \\
\hat{A}_{{\textit {\tiny TE}}}^{b}\\
\hat{B}_{{\textit {\tiny TE}}}^{f} 
\end{array}
\right)=
\left(
\begin{array}{cccc}
T_{11} & T_{12} & T_{13} &  T_{14} \\
T_{21} & T_{22} & T_{23} & T_{24} \\
T_{31} & T_{32} & T_{33} & T_{34} \\
T_{41} & T_{42} & T_{43} & T_{44} \\
\end{array}
\right)
\left(
\begin{array}{c}
\hat{A}_{{\textit {\tiny TM}}}^{f} \\
\hat{B}_{{\textit {\tiny TM}}}^{b} \\
\hat{A}_{{\textit {\tiny TE}}}^{f}\\
\hat{B}_{{\textit {\tiny TE}}}^{b} 
\end{array}
\right) \label{Tmatrix} \ee
where $\hat{A}_{r}^{f,b}:=(\hat{A}_{r,0}^{f,b}, \ldots, \hat{A}_{r,N+1}^{f,b})^T$ and similarly for $\hat{B}_{r}^{f,b}$. Here $\hat{A}_{{\textit {\tiny TM}},0}^f$, $\hat{A}_{{\textit {\tiny TM}},0}^b$, $\hat{B}_{{\textit {\tiny TM}},0}^f$ and $\hat{B}_{{\textit {\tiny TM}},0}^b$ represent the forward and backward propagating SPP excitation operators, as specified in Eq.~(\ref{SPPfromTM}). To obtain the full transfer matrix $T$, one must calculate the elements of the $D$ and $F$ matrices, which consist of the various coupling coefficients $ {\cal C}^{ji}_{\mu\nu}({\bf k}_j,q_j;{\bf k}_i,q_i)$ (see Appendix B). The matrices $D$ and $F$ then enter into the elements of $T$ as outlined in Appendix D.

One of the fundamental differences between the scattering of SPPs described here and the scattering of photons at a standard optical interface is that the spatial profile of an incoming SPP's wavefunction is modified due to the variation in the permitivitty of the materials on either side of the interface (metal and dielectric). As a result, the reflection and refraction of SPPs do not follow the standard laws of optics for determining the various scattering angles involved, such as Snell's law and the usual Fresnel equations~\cite{Steg,Elser}. Fortunately, all the information required to obtain the scattering angles for the SPPs and radiation excitations is contained within the coupling elements. In particular, one can immediately and in a straightforward manner calculate the angles for all the SPPs involved in the scattering process. 

As can be seen from Eq.~(\ref{compactcouple}) for the coupling coefficients $ {\cal C}^{ji}_{\mu\nu}({\bf k}_j,q_j;{\bf k}_i,q_i)$, the magnitude of the wavevector in the $\hat{\bf y}$ direction, $k_y$, and frequency of the excitation, $\omega$, must be the same on either side of the interface. This allows one to immediately specify reflected and transmitted angles for all SPP excitations involved. This can be achieved because for SPPs, once $\omega$ is set, the DR given by $k=(\omega/c)(\epsilon_{d}\epsilon_{m}/(\epsilon_{d}+\epsilon_{m}))^{1/2}$ automatically sets the value of $k$, and with $k_y$ also set one directly obtains the value of $k_x$. For instance, consider a single-SPP excitation in region $i$ incoming at specific angle $\theta_{i_i}$ with a set frequency $\omega_i$. Recently it has been shown that such a quantum excitation can be efficiently generated on the metal-dielectric interface using an attenuated-reflection geometry~\cite{TSPP,DSPP}. With all the material properties of the interface configuration set, using the DR for SPPs in region $i$, the frequency and angle values result in a corresponding wavenumber $k_i$ and wavevector ${\bf k}_i=k_{x,i}\hat{\bf x}+k_{y,i}\hat{\bf y}$. From this, we can calculate the wavevector for the transmitted SPP in region $j$, as the condition $\omega_j=\omega_i$ and $k_{y,j}=k_{y,i}$ must be met across the interface. From the wavevector, the transmitted angle $\theta_{i_t}$ can be found. Note that due to the monoticity of the DR, there is only one SPP in region $j$ that can be excited by a single SPP incident in region $i$. Furthermore, based on the above arguments, it is straightforward to check that the incident angle is equal to the reflected angle, {\it i.e.}~$\theta_{i_i}=\theta_{i_r}$. For the radiation excitations such a direct specification of the angles, as in the SPP case, is not possible. This is because $k$ (and therefore $k_x$) depends also on the value of the parameter $q$. Thus, for a set $\omega$ and $k_y$ there are a continuum of radiation excitations which couple to an incoming SPP, each one exiting the scattering region at a particular angle with a given probability amplitude defined by the transfer matrix $T$.

\section{Beamsplitter characterization}
A beamsplitter is one of the most important devices in linear optics. It constitutes an essential building block in any classical optical setup~\cite{ST} and in the context of QIP it plays a crucial role in optical implementations of quantum gates~\cite{NC,KLM}. In the simplest case of a lossless optical device, the action of a photonic beamsplitter can be described mathematically by a 2$\times$2 matrix that relates the two output ports to the two input ports. It is possible to derive further constraints and relations for the elements of this matrix, the reflection and transmission coefficients, based on simple energy conservation arguments or on the fact that the device is passive, reciprocal, and lossless \cite{OM,Loudon}. 

The situation is somewhat more involved for SPPs because the incoming SPP excitations are not only coupled to the outgoing SPP ones, but also to a continuum of radiation excitations. In this sense, the problem is similar to the case of a lossy beamsplitter and, with some care taken - using the theory derived in the previous section - it can be treated as such. Furthermore, we note that the geometry considered here consists of two different media on either side of the interface, thus unlike the standard photonic beamsplitter scenario, one must also take into account the different group velocities associated with the propagating plasmonic excitations. For the present discussion we will be interested in the main beamsplitting properties in the immediate vicinity of the interface. Propagation issues will be discussed in the next section. 

It is also important to note that while a beamsplitter is conventionally defined as a device used to divide a single incident beam (on one side) into two beams (one on either side) with a ratio of intensities, in this work we focus on the description of a beamsplitting device where SPPs can be incident on the interface from one side or the other, and even from both sides at the same time, but with the condition that the output modes for the excitations are the same, regardless of which side the SPPs are incident on. This scenario is the usual symmetric case considered for a beamsplitter operating at the quantum level which mixes two input fields and produces two output fields~\cite{OM,Loudon}. In order to highlight that the device must behave in this symmetric way we focus our discussion on the latter case - when a single-SPP is incident on both sides. One can easily modify this for other cases, such as when a single-SPP is incident from only one side or the other by placing one of the SPP excitation field modes in the vacuum state~\cite{Loudon}. 

First, consider that an SPP is incoming from region $j$ at an angle $\theta_{i_j}$ and another is incoming from region $i$ at an angle $\theta_{i_i}$. In the present context, for a quantum beamsplitter operating faithfully at the single-SPP level it is essential that the input SPP excitations produce outputs which are indistinguishable in both their spatial and temporal degrees of freedom, regardless of the transmission and reflection coefficients. One major requirement for spatial indistinguishability is that the angle $\theta_{j_t}$, corresponding to the transmitted SPP into region $i$, must be equal to the angle $\theta_{i_r}$ of the reflected SPP excitation incoming from region $i$. In addition, we must ensure that $\theta_{j_r}=\theta_{i_t}$. These conditions are shown more clearly in Fig.~\ref{fig2}~{\bf (b)}. Surprisingly, such a matching set of splitting angles can be achieved. By setting the incidence angle for the incoming SPP in region $j$ to $\theta_{j_i}=\theta_{i_t}=\cos^{-1}[(1-\epsilon_{d,i}\epsilon_{m,i}(\epsilon_{d,j}+\epsilon_{m,j})\sin^2 \theta_{i_i}/\epsilon_{d,j}(\epsilon_{d,i}+\epsilon_{m,i})\epsilon_{m,j})^{1/2}]$, we find that the transmitted angle $\theta_{j_t}=\theta_{i_r}$ regardless of the frequency $\omega$ of the excitation. Thus one is always able to achieve the required matching for the angles in the scattering process, as long as $\theta_{i_i}$ results in $\theta_{i_t}\in [0,\pi/2]$, otherwise plasmonic total internal reflection (TIR) will occur. This symmetric splitting regime has not been investigated before, even in the classical domain. In what follows in our beamsplitter characterization it should always be assumed that the matching angle set has been chosen as specified above, regardless of the parameter regime being studied. All angles will be specified where relevant.

For temporal indistinguishability, it is important that the incoming SPPs have exactly the same arrival time at the interface, as well as having the same spectral profile (bandwidth). For the former, we assume this can always be achieved experimentally, constituting a practical issue rather than anything fundamental in the theory. For the latter, we will begin our discussion with a simple single-mode picture and discuss wavepacket excitations later. 

Continuing with our aim to make the scattering interface work as an SPP beamsplitter, we must also quantify how well SPP excitations couple to each other and not to the radiation excitations. From the transfer matrix given in Eq.~(\ref{Tmatrix}) one can link the output to the input operators for the SPPs using the relations
\bea
\hat{a}^{b}({\bm \omega}_i) &=&(T_{11})_{00}\,\hat{a}^{f}({\bm \omega}_i) +\sum_{\ell=1}^{N+1}(T_{11})_{0\ell} \, \hat{A}_{{\textit {\tiny TM}},\ell}^{f}({\bm \omega}_i,q_i)   \label{outin1}  \\
&&  \hskip0cm+   (T_{12})_{00}\,\hat{b}^{b}({\bm \omega}_j) +\sum_{\ell=1}^{N+1}(T_{12})_{0\ell}\hat{B}_{{\textit {\tiny TM}},\ell}^{b}({\bm \omega}_j,q_j)   \nonumber \\
&& \hskip0cm +  \sum_{\ell=1}^{N+1} (T_{13})_{0\ell}\,\hat{A}_{{\textit {\tiny TE}},\ell}^{f}({\bm \omega}_i,q_i) +\sum_{\ell=1}^{N+1}(T_{14})_{0\ell}\hat{B}_{{\textit {\tiny TE}},\ell}^{b}({\bm \omega}_j,q_j),  \nonumber \\
\hat{b}^{f}({\bm \omega}_j) &=&(T_{21})_{00}\,\hat{a}^{f}({\bm \omega}_i) +\sum_{\ell=1}^{N+1}(T_{21})_{0\ell}\hat{A}_{{\textit {\tiny TM}},\ell}^{f}({\bm \omega}_i,q_i)    \label{outin2}  \\
&& \hskip0cm+     (T_{22})_{00}\,\hat{b}^{b}({\bm \omega}_j) +\sum_{\ell=1}^{N+1}(T_{22})_{0\ell}\hat{B}_{{\textit {\tiny TM}},\ell}^{b}({\bm \omega}_j,q_j)    \nonumber  \\
&&  \hskip0cm+  \sum_{\ell=1}^{N+1} (T_{23})_{0\ell}\,\hat{A}_{{\textit {\tiny TE}},\ell}^{f}({\bm \omega}_i,q_i) +\sum_{\ell=1}^{N+1}(T_{24})_{0\ell}\hat{B}_{{\textit {\tiny TE}},\ell}^{b}({\bm \omega}_j,q_j)  \nonumber  .
\eea
Similar relations can be written for the radiation excitations. 

In order to achieve perfect matching of the quantized electromagnetic field at the interface, we have included evanescent radiation excitations in our theory and therefore increased $q_{\rm cut}$ from $\omega_i (\epsilon_{d,i})^{1/2}/c$ to $\omega_i (\epsilon_{d,i}-\epsilon_{m,i})^{1/2}/c$.. These excitations represent a contribution to the total field which decays exponentially from the interface in the $\hat{\bf x}$-direction (in addition to the usual decay in the $\hat{\bf z}$-direction parameterized by the variable $q$, as described in section II A). Although these excitations do not propagate they are important in the study of near-field phenomena. For instance, such near-field radiation could be exploited to achieve coupling with a resonant auxiliary system placed in close proximity to the interface. However, by extending the quantization procedure followed for the propagating radiation (see Section II) to decaying evanescent radiation, we have that the photon number operator of the evanescent excitations is not Hermitian. This implies that such an operator is not a physical observable. Nonetheless, when we look at the scattering problem within the far-field approximation we have that the transfer matrix $T$ is indeed unitary and provides amplitudes for the transfer of the physically well-defined propagating excitations. A more rigorous treatment of the evanescent excitations along the lines of Ref. \cite{CM} is possible, but such a treatment is beyond of the scope of this work. Moreover, the appearance of evanescent excitations and their non-reciprocal behavior on either side of the interface leads to a general lack of reciprocity in the splitting device, the effects of which can be measured, as described next.

\begin{figure}[t]
\centerline{\psfig{figure=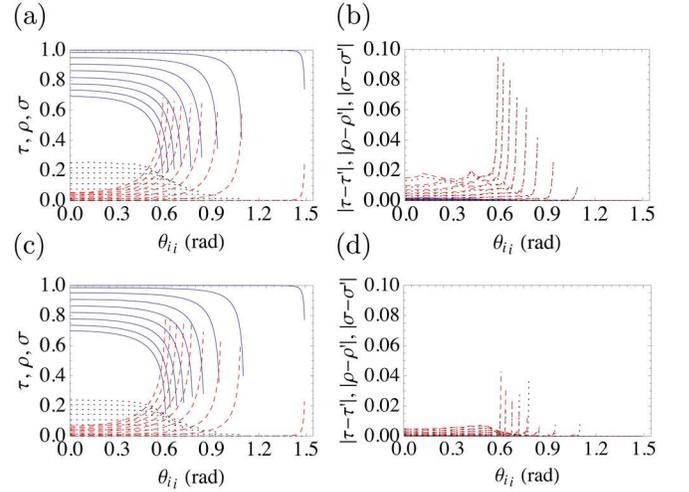,width=9cm}}
\caption{(Color online) Scattering interface as an SPP beamsplitter for $\lambda_0=790$nm shown in panels {\bf (a)}~\&~{\bf (b)} and for $1500$nm in panels {\bf (c)}~\&~{\bf (d)}. {\bf (a)}~\&~{\bf (c)}: Forward SPP transmission $\tau$ (solid lines), reflection $\rho$ (dashed lines) and scattering $\sigma$ (dotted lines) coefficients versus the incidence angle $\theta_{i_i}$. In all plots $\epsilon_{d,i}$ increases in steps of 0.25 from 1 to 3 going from right to left (bottom to top for the scattering coefficient). {\bf (b)}~\&~{\bf (d)}: Coefficients $|\tau-\tau'|$, $|\rho-\rho'|$, $|\sigma-\sigma'|$. In all cases, $\epsilon_{d,j}=1$ and $\omega_{p,{i/j}}=\omega_p^{silver}=1.402\times 10^{16}$rad/s are chosen. Note that in the plots of panels {\bf (a)}~\&~{\bf (c)} the coefficients $\tau$, $\rho$ and $\sigma$ are limited at large angles of $\theta_{i_i}$. This is due to a reduction in numerical precision close to plasmon TIR for which we do not include the corresponding values.}
\label{fig3}
\end{figure}

From Eqs.~(\ref{outin1})~and~(\ref{outin2}) we define the coefficients for the SPP beamsplitter of an incoming SPP in region $i$ as follows, transmission $\tau=|(T_{21})_{00}|^2$, reflection $\rho=|(T_{11})_{00}|^2$, and total loss (into propagating radiation) $\sigma=\sum_{\ell=1}^{mmax_i+1}|(T_{11})_{\ell0}|^2 +\sum_{\ell=1}^{mmax_j+1} |(T_{21})_{\ell0}|^2   + \sum_{\ell=1}^{mmax_i+1}  |(T_{31})_{\ell0}|^2 +  \sum_{\ell=1}^{mmax_j+1} |(T_{41})_{\ell0}|^2 $, where $mmax_i$  ($mmax_j$) stands for the number of non-evanescent (propagating) radiation excitations in region $i$ ($j$) upon discretization. Similarly we define the coefficients of an incoming SPP in region $j$ as $\tau'=|(T_{21})_{00}|^2$ for transmission, $\rho'=|(T_{11})_{00}|^2$ for reflection, and $\sigma'=\sum_{\ell=1}^{mmax_j+1}|(T_{11})_{\ell0}|^2 +\sum_{\ell=1}^{mmax_i+1} |(T_{21})_{\ell0}|^2   + \sum_{\ell=1}^{mmax_j+1}  |(T_{31})_{\ell0}|^2 +  \sum_{\ell=1}^{mmax_i+1} |(T_{41})_{\ell0}|^2 $ for total loss, which are calculated for the reverse configuration (from $j$ to $i$). Thus, using the above relations we can rearrange the transfer matrix Eq. (\ref{Tmatrix}) as
\be 
\left(
\begin{array}{c}
\hat{a}^{b}({\bm \omega}_i) \\
\hat{b}^{f}({\bm \omega}_j) \\
\vdots
\end{array}
\right)=
\left(
\begin{array}{ccc}
e^{i \varphi} \sqrt{\rho} & \sqrt{\tau'} & \hdots \\
\sqrt{\tau} & -e^{-i \varphi} \sqrt{\rho'} & \hdots \\
\vdots & \vdots & \ddots
\end{array}
\right)
\left(
\begin{array}{c}
\hat{a}^{f}({\bm \omega}_i) \\
\hat{b}^{b}({\bm \omega}_j) \\
\vdots
\end{array}
\right), \label{BSmatrix} \ee
where the phases of the coefficients stem automatically from the matching of the transverse components of the wavefunctions in Eqs.~(\ref{transa})~and~(\ref{transb}). We find that when $\epsilon_{d,i}>\epsilon_{d,j}$ the phase $\varphi=0$ and when $\epsilon_{d,i}<\epsilon_{d,j}$ the phase $\varphi=\pi$. In order to check numerically how close the interface devised here is to a reciprocal one, we require the conditions $\tau\simeq \tau'$, $\rho\simeq \rho'$ and $\sigma\simeq \sigma'$ to be met~\cite{Loudon}.  

\begin{figure}[b]
\centerline{\psfig{figure=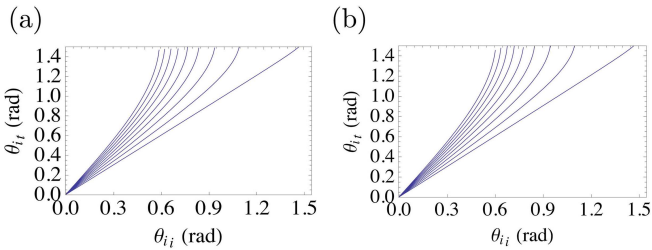,width=9cm}}
\caption{(Color online) Transmission angle $\theta_{i_t}$ as a function of the incidence angle $\theta_{i_i}$. {\bf (a)}: $\lambda_0=790$nm. {\bf (b)}: $\lambda_0=1500$nm. In both plots $\epsilon_{d,i}$ increases in steps of 0.25 from 1 to 3 going from right to left.}
\label{figtrans}
\end{figure}

\begin{figure*}[t]
\centerline{\psfig{figure=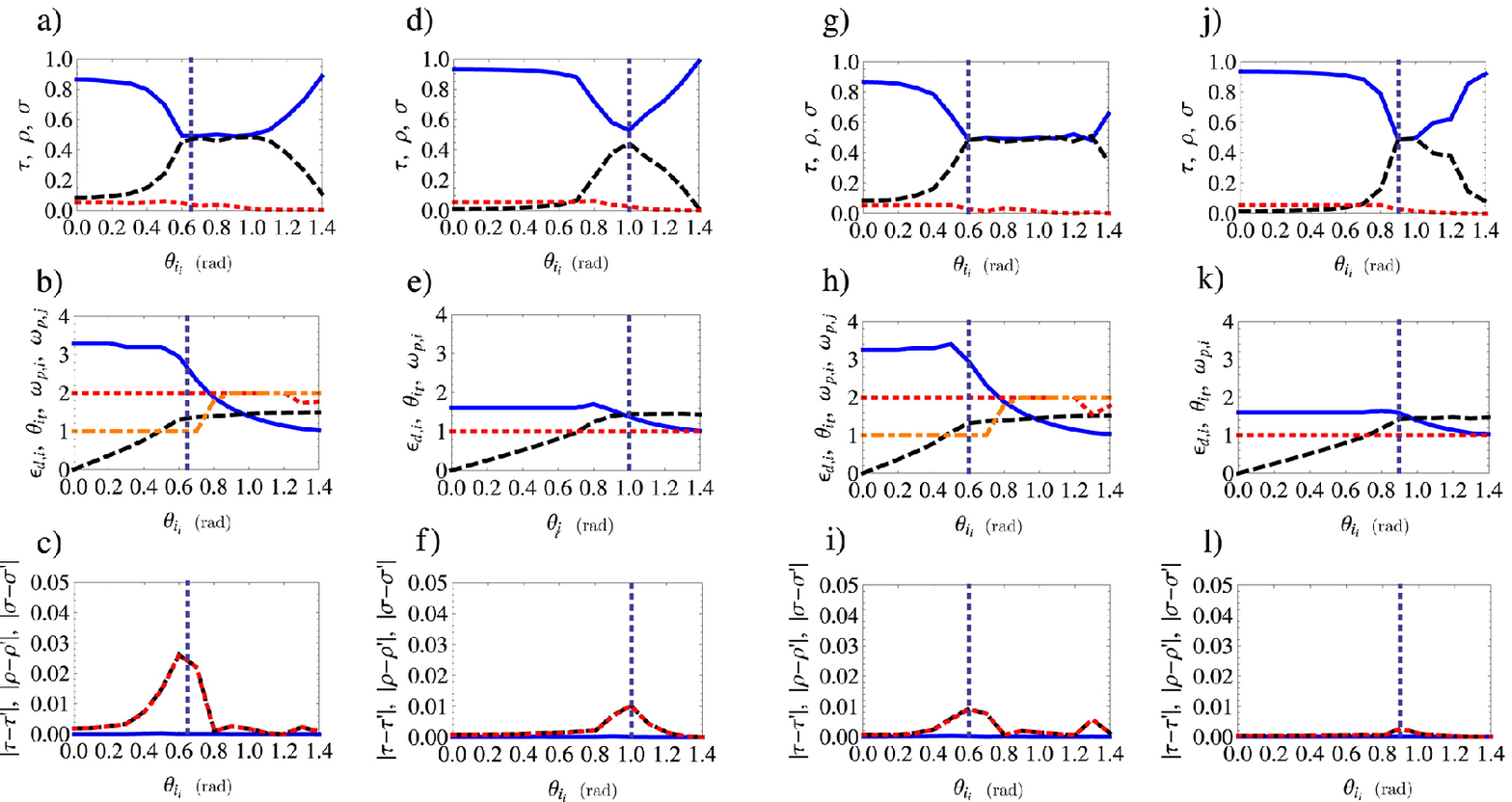,width=17.0cm}}
\caption{(Color online) Results of the optimization procedure for achieving a 50:50 SPP beamsplitter with the scattering interface over a range of incidence angles $\theta_{i_i}$ for $\lambda_0=790$nm ($1500$nm). {\bf (a)} ({\bf (g)}): SPP transmission $\tau$ (solid), reflection $\rho$ (dash), and scattering $\sigma$ (dotted) coefficients. {\bf (b)} ({\bf (h)}): Values of $\epsilon_{d,i}$ (solid), $\theta_{i_t}$ (dash), $\omega_{p,i}$ (dotted), $\omega_{p,j}$ (dash-dot). {\bf (c)} ({\bf (i)}): Coefficients $|\tau-\tau'|$ (solid), $|\rho-\rho'|$ (dash), $|\sigma-\sigma'|$ (dotted). {\bf (d-f)} ({\bf (j-l)}): Same as {\bf (a-c)} ({\bf (g-i)}) but with the restriction $\omega_{p,i}=\omega_{p,j}=1$. In all cases, $\epsilon_{d,j}=1$ is chosen and values of $\omega_{p,{i/j}}$ are measured in units of $\omega_p^{silver} =1.402\times 10^{16}$rad/s. The vertical dotted lines designate optimized angles and the corresponding material parameters for the 50:50 beamsplitter.}
\label{fig5}
\end{figure*}

In Fig.~\ref{fig3} we show how the SPP transmission, reflection and scattering coefficients vary as a function of the incidence angle $\theta_{i_i}$ and the value of $\epsilon_{d,i}$ for free-space wavelengths $\lambda_0=790$nm {\bf (a)}~\&~{\bf (b)} and $1500$nm {\bf (c)}~\&~{\bf (d)} ($\omega_i=2 \pi c/\lambda_0$). These wavelengths have been chosen as examples to illustrate our results as they correspond to those used regularly in quantum optics experiments~\cite{ST}. The range of $\epsilon_{d,i}:1 \to 3$ (in steps of 0.25 going from right to left) chosen corresponds to that which can be obtained with the use of basic optical materials~\cite{Boyd}. Here, as an example we set $\epsilon_{d,j}=1$, with the metal in both regions modeled as silver. We also choose the number of radiation excitations to be $N+1=200$ and $q_{\rm cut}=10\,{\rm max}[(\epsilon_{d,i}k_i^2-k^2_{y,i})^{1/2},(\epsilon_{d,j}k_j^2-k^2_{y,j})^{1/2}]$ in order to satisfy convergent behavior of the coefficients~\cite{Oult}. One can see from panels {\bf (a)}~and~{\bf (c)} that, in the case of normal incidence $\theta_{i_i}=0$, a large value of $\epsilon_{d,i}\simeq3$ is required in order to observe a significant departure of the transmission coefficient $\tau$ from unity. However, this is accompanied by a significant increase in the value of $\sigma$  corresponding to scattering into radiation excitations. To avoid such a scenario one could use lower $\epsilon_{d,i}$'s and vary the angle $\theta_{i_i}$ instead. However, one must be careful in doing so for two reasons. First, there exists a critical value of the incidence angle for which plasmonic TIR takes place. In order to ensure that we do not reach the plasmon critical angle our computation stops at those values of $\theta_{i_i}$ near which TIR for the radiation excitations occurs. This can be seen more clearly in Fig.~\ref{figtrans}, where we show the dependence of the transmission angle $\theta_{i_t}$ (as it approaches $\pi/2$ rad) as a function of the incident angle $\theta_{i_i}$ and $\epsilon_{d,i}$. However, it is the stable region around this critical angle that there is a wide range of opportunity in setting the transmission and reflection coefficients. The second reason one needs to be careful is directly related to the first, and it is that there is a rapid change in the transmission and reflection coefficients in the region of opportunity, as shown in Fig.~\ref{fig3}~{\bf (a)}~and~{\bf (c)}. Thus, the coefficients are quite sensitive to the incidence angle $\theta_{i_i}$. In addition, Fig.~\ref{fig3}~{\bf (b)} and {\bf (d)} one can see that the difference between the coefficients, i.e. $|\tau-\tau'|$, $|\rho-\rho'|$, $|\sigma-\sigma'|$, is relatively small with varying $\theta_{i_i}$ and $\epsilon_{d,i}$.

In our calculations we have set $\epsilon_{d,i}=1$ for simplicity. In general, one needs to be careful with the opposite configuration, where the incoming excitation goes from a higher permittivity region to a lower one. This is because the coupling to evanescent radiation appears to be more substantial. We also note that we have verified numerically that when metals with highly negative permittivity are used (challenging from a fabrication viewpoint~\cite{Pendry,WasserShan}), compared to the values of the dielectrics, it is possible to favor the stability of the SPP excitations over the radiation, suppressing losses into radiation even more.

While we have considered the dielectric media and metal in both regions to be passive (fixed) in this work, our theory applies equally well to the case of active materials, such as electro-optical polymers or voltage programmable liquid~\cite{Brown} for the dielectric media, and metal-semiconductor quantum well structures~\cite{Ambati} in place of the metal. The use of such active material would provide greater flexibility and external control of the parameter regime of the constructed beamsplitter, and thus the range of transmission and reflection coefficients, without the need for re-fabrication of the device. Moreover, such `on-chip' beamsplitters may be created and destroyed at desired locations using the correct associated electrical circuitry~\cite{Elser}.

In summary, we have found that operating close to the critical angle for $\theta_{i_i}$ provides great flexibility in obtaining a range of values of the transmission and reflection coefficients, however it comes at the expense of the reciprocity of the beamsplitter, loss into radiation, as well as the necessity to accurately set the angle $\theta_{i_i}$ due to the rapidly changing values of the coefficients. As is evident from the above initial investigation, the physical properties of the materials involved together with the incidence angle determine how well one can optimize an SPP beamsplitter to provide a desired splitting ratio. 

While we have found that a range of splitting ratios can be reached, as an example we now fix the goal of achieving a 50:50 plasmonic beamsplitter, a versatile component in classical and quantum optics~\cite{ST}. By allowing the different parameters of the interface to vary, such as the dielectric material, metal, frequency and angle of incidence, we seek to optimize the beamsplitter's performance.
\begin{figure}[b]
\centerline{\psfig{figure=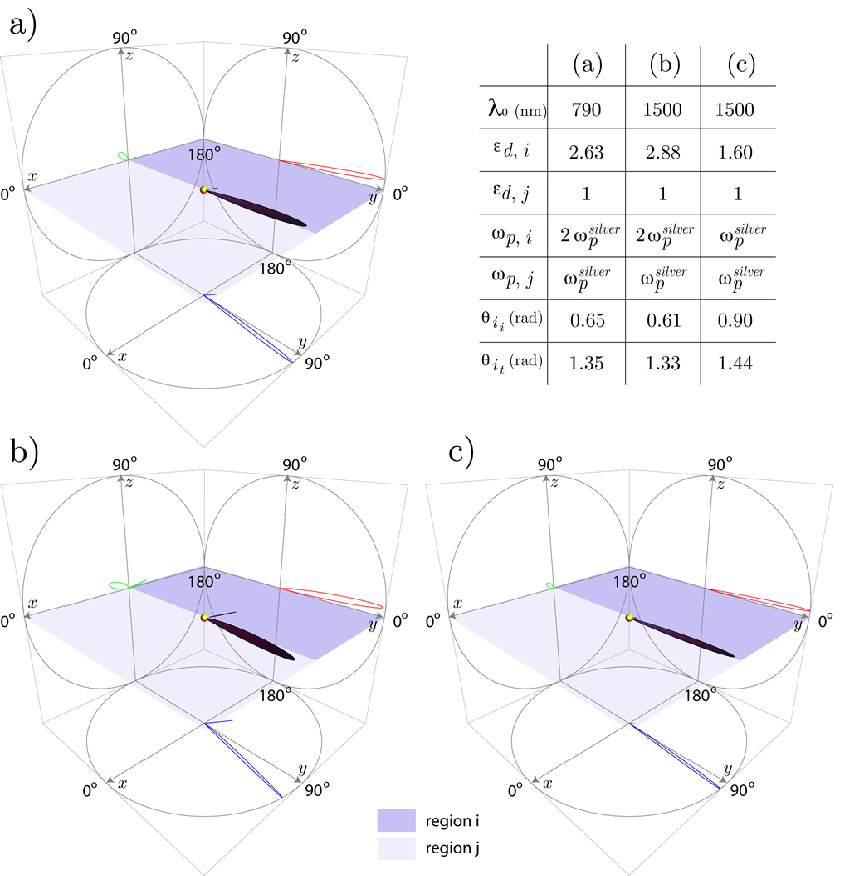,width=8.7cm}}
\caption{(Color online) Scattered TM radiation plots resulting from the scattering of an incoming SPP from region $i$. {\bf (a)}: 50:50 beamsplitter for $\lambda_0=790$nm. Here, the parameters chosen correspond to those defined by the dotted line of Fig.~\ref{fig5}~{\bf (a)}, where the metal dielectric is allowed to vary. {\bf (b)}: 50:50 beamsplitter for $\lambda_0=1500$nm. Here, the parameters chosen correspond to those defined by the dotted line of Fig.~\ref{fig5}~{\bf (g)}. {\bf (c)}: 50:50 beamsplitter for $\lambda_0=1500$nm. Here, the parameters chosen correspond to those defined by the dotted line of Fig.~\ref{fig5}~{\bf (j)}, where the metal is the same on both sides of the interface. As one cannot reach 50:50 for $\lambda_0=790$nm in this regime, such a plot has not been included.}
\label{fig7}
\end{figure}

In Fig.~\ref{fig5} we show the numerical results for an optimization procedure we have followed in order to bring the scattering configuration as close as possible to a 50:50 beamsplitter for $\lambda_0=790$nm {\bf (a-f)} and $1500$nm {\bf (g-l)}. The intent of these plots is to show that as the angle $\theta_{i_i}$ is varied, the dielectric constants and plasma frequencies of the metals on either side can be changed in order to bring the beamsplitter as close as possible to a 50:50 splitting ratio. Here we have imposed several constraints on the acceptable configuration. First, the maximum amount of scattering into radiation $\sigma$ should be 5\% or less and second, in order to force the interface to work as a reciprocal device, we constrain the differences $|\tau-\tau'|$, $|\rho-\rho'|$ and $|\sigma-\sigma'|$ to be lower than 2.5\%. One can clearly see in Fig.~\ref{fig5}~{\bf (a)} ({\bf (g)}) that a 50:50 beamsplitter can be reached for $\lambda_0=790$nm ($\lambda_0=1500$nm) if we allow the metals to vary across the interface as shown in panel {\bf (b)} ({\bf (h)}), assuming $\omega_p^{silver}  \leq \omega_{p,{i/j}} \leq 2 \omega_p^{silver}$, for a range of incidence angles $\theta_{i_i}$. Such a range in the value of the plasma frequency for the metals should be possible, for instance by embedding highly conducting surfaces perforated by holes~\cite{Pendry} or depositing patterned metallic films on semiconductor structures~\cite{WasserShan}. On the other hand, if we set the metal as silver on both sides of the interface as a less demanding scenario as shown in panel {\bf (e)} and {\bf (k)} for $\lambda_0=790$nm and $\lambda_0=1500$nm respectively, we can reach a 50:50 beamsplitter, although only at the higher wavelength of $\lambda_0=1500$nm, as highlighted by panels {\bf (d)} and {\bf (j)}.

Throughout the above analysis we have assumed that the scattering of SPP excitations into radiation excitations can be treated as a loss mechanism. In order to ensure that this is indeed the case and that the radiation excitations are not able to interfere with any SPP detection process or subsequent SPP beamsplitter operations one must check that the radiation excitations scatter in a direction such that their quantum degrees of freedom can effectively be discarded, or `traced out'~\cite{NC}, from the SPP dynamics. In Fig.~\ref{fig7} we show the scattering angles of the resulting TM radiation excitations from an incoming SPP excitation in region $i$ with $\lambda_0=790$nm ({\bf (a)}) and $1500$nm ({\bf (b)} and {\bf (c)}). The parameters chosen for the materials and the incidence angles correspond to those defined in the caption by the respective dotted lines of Fig.~\ref{fig5} and given in the table of Fig.~\ref{fig7}. For these three scenarios, the scattered TE radiation is an order of magnitude less than the TM radiation. As the total radiated power has a maximum of 5\%, we focus on the TM excitations as the main sources of power loss. The plots of Fig.~\ref{fig7} essentially show the direction of the $2(N+1)$ scattered radiation excitations - defined by the wavevectors $(k_{x,i},k_{y,i},q_{i})$ and similarly for region $j$ - multiplied by their contribution to the total scattered power (normalized by the largest contributor). The start (end) points of these normalized `power-vectors' lie at the origin (on the curves). For both $\lambda_0=790$nm and $1500$nm one can clearly see that the resulting radiation excited in both forward and backward directions exits the scattering region at reasonably large enough angles from the $x$-$y$ plane such that it is possible to trace them out from the system dynamics and thus treat the radiation as a truly lossy mechanism. Similar plots can be made for the SPP incoming from region $j$ with the same parameters and resulting in the same conclusions. Note the fact that we can trace-out the radiation excitations is not a general rule even though it applies to the parameter range optimized and investigated here. One should analyse the TM and TE radiation scattering on a case-by-case basis.

\section{Quantum interference}   
In our derivation of the properties of the scattering interface, we have worked explicitly with the quantized form of SPP and photon radiation fields. We have done this because at the quantum level there are interactions between SPP states which cannot be described within a classical framework. In this context a particularly enlightening scenario is the interference of two single-SPPs. We now apply our theory to this case as an explicit example of the necessity of a quantum theory for plasmonic scattering.
\begin{figure*}[t]
\centerline{\psfig{figure=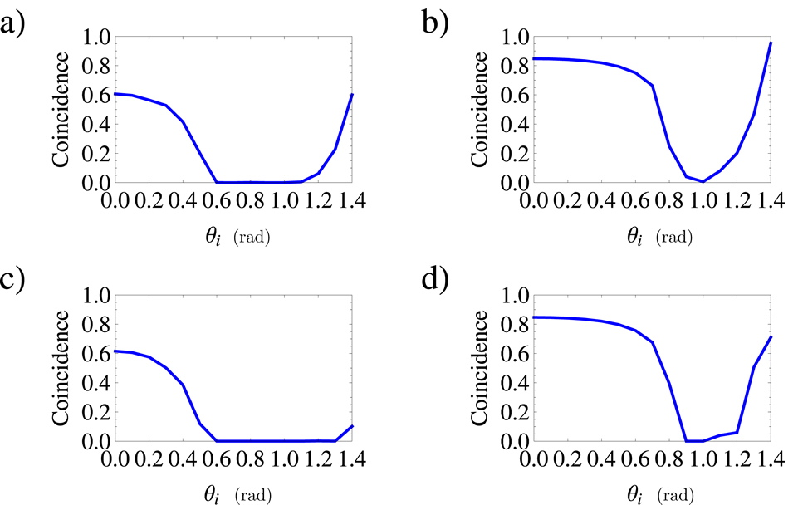,width=9.cm}\qquad \qquad  \psfig{figure=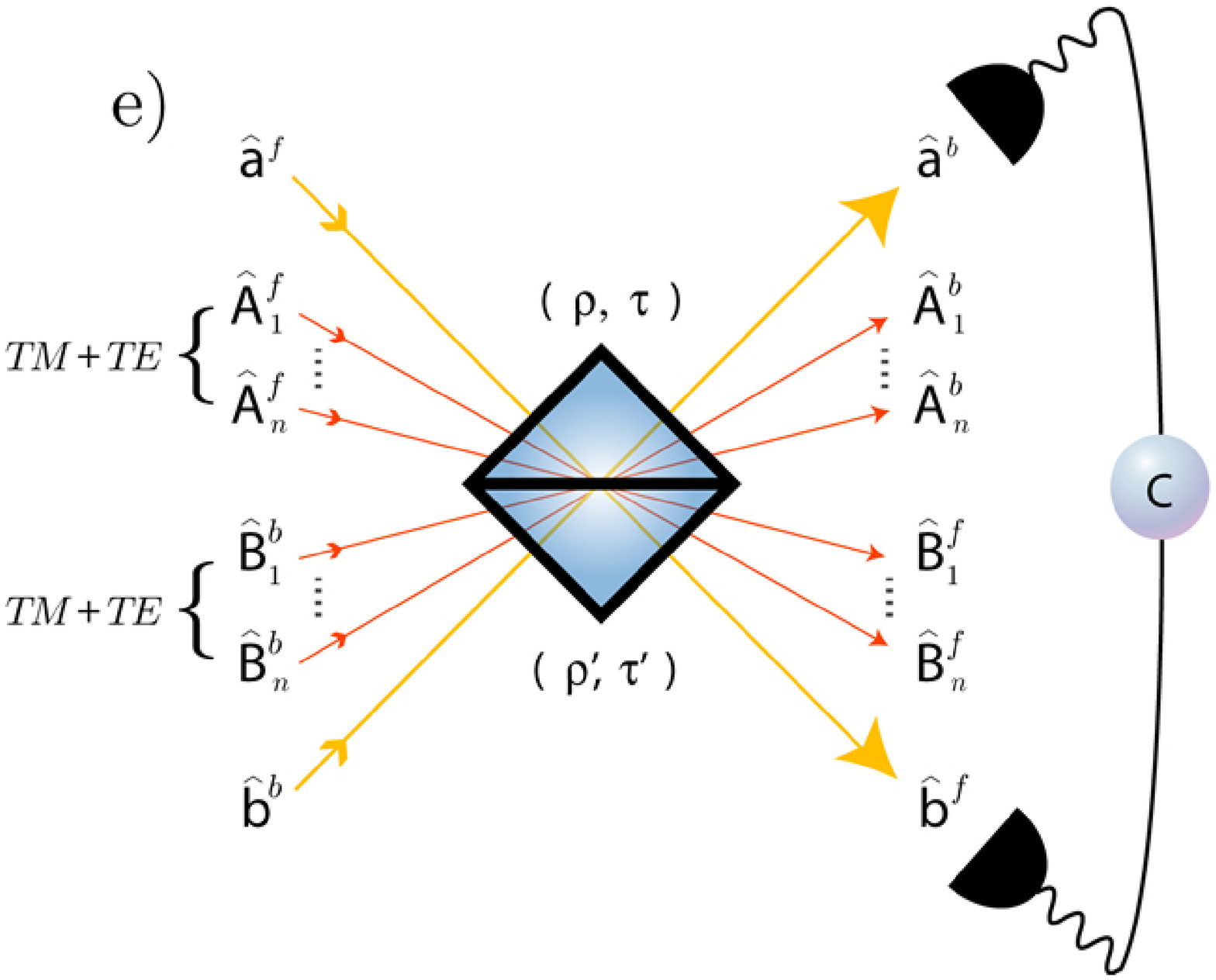,width=6.5cm}}
\caption{(Color online) First-order quantum interference of SPPs where the parameters chosen correspond to the optimization procedure used for the 50:50 SPP beamsplitter described in Fig.~\ref{fig5} for $\lambda_0=790$nm ($1500$nm). {\bf (a)} ({\bf (c)}): Two-fold coincidence probabilities. {\bf (b)} ({\bf (d)}): Two-fold coincidence probabilities with the restriction $\omega_{p_i}=\omega_{p_j}=\omega_p^{silver}$. In all cases, $\epsilon_{d_j}=1$ is chosen. {\bf (e)}: Optical beamsplitter analogy for the plasmonic beamsplitter with two detectors measuring the presence of an SPP at both ports at the same time ($C$ represents a coincidence detection circuit).}
\label{fig9}
\end{figure*}

Consider an input state with two SPPs, one on each side of the interface. For simplicity, we assume that both SPPs are monochromatic, with the same frequency, and arrive in coincidence. Thus temporal indistinguishability is immediately satisfied. Although, more realistic Gaussian wavepackets will be considered in a subsequent study with more complex scattering geometries~\cite{Ballester}, in the present work we have verified numerically that when a 10nm-width wavepacket centered at the frequencies corresponding to $\lambda_0=$ 790nm and 1500nm are considered, the values of the coefficients of the scattering matrix do not change appreciably. Thus a single-mode picture can be adopted as a close approximation for this initial investigation. We also note that the spectral shape of a wavepacket does not change on passing from one region to another. The situation changes during propagation, where after a given distance, the group velocity related to two different media causes a deformation (broadening) of the wavepacket at different rates~\cite{ST}. This can be compensated with the incorporation of appropriate dielectic media on the metal surface~\cite{Ballester} or by considering only short propagation distances. 

We write $\ket{1,{\bf 0}}_{p_i r_i} \ket{1,{\bf 0}}_{p_j r_j}$  in order to represent the input state in the Schr\"odinger picture, which consists of a single SPP on each side of the interface plus vacuum for the input radiation. To calculate the output after scattering occurs at the interface we apply the transfer matrix in Eq.~(\ref{BSmatrix}) to the input state, written in the Heisenberg picture as
\begin{equation}
 \ket{1,{\bf 0}}_{p_i r_i} \ket{1,{\bf 0}}_{p_j r_j} =  \hat{a}^{\dag f}({\bm \omega}_i) \hat{b}^{\dag b}({\bm \omega}_j)  \ket{0,{\bf 0}}_{p_i r_i} \ket{0,{\bf 0}}_{p_j r_j}.  \label{instate}
\end{equation}
Due to the non-zero value of the couplings between SPP and photon radiation, the output state will contain terms in which the radiation excitations are populated with a given probability amplitude. More precisely, for the input defined in Eq.~(\ref{instate}) we have the output
\begin{eqnarray}
&& \hskip-0.6cm  \left( e^{i \varphi} \sqrt{\rho} \, \hat{a}^{\dag b}({\bm \omega}_i) +  \sqrt{\tau} \, \hat{b}^{\dag f}({\bm \omega}_j) + \ldots \right)\times    \nonumber \\
&&  \hskip-0.2cm  \left( \sqrt{\tau} \, \hat{a}^{\dag b}({\bm \omega}_i) - e^{-i \varphi} \sqrt{\rho} \,  \hat{b}^{\dag f}({\bm \omega}_j) + \ldots \right)   \ket{0,{\bf 0}}_{p_i r_i} \ket{0,{\bf 0}}_{p_j r_j}  \nonumber \\
&&\hskip-0.5cm = \sqrt{2 \rho \tau}  \left(  e^{i \varphi}   \ket{2,{\bf 0}}_{p_i r_i} \ket{0,{\bf 0}}_{p_j r_j}  - e^{-i \varphi}   \ket{0,{\bf 0}}_{p_i r_i} \ket{2,{\bf 0}}_{p_j r_j}  \right)\nonumber \\
&&\hskip0.9cm+  \left( \tau - \rho \right)  \ket{1,{\bf 0}}_{p_i r_i}  \ket{1,{\bf 0}}_{p_j r_j}   + \ldots      \, , \label{state}
\end{eqnarray}
where the dots in the last line account only for terms in the superposition with strictly zero SPP excitations. Assuming that the state at the output SPP ports after the scattering process can be measured ideally~\cite{idealmeas}, then one can calculate the probability to measure a single SPP at each output port in `coincidence', {\it i.e.} a joint detection event. This probability is found to be $P_{coincidence}=\left( \tau - \rho \right)^2$~\cite{Loudon}. In particular, for a 50:50 beamsplitter, a zero probability is obtained. This is a result of the well-known Hong-Ou-Mandel effect~\cite{HOM} and it is useful in characterizing how well quantum interference can occur for a particular scattering device. This effect is purely quantum-mechanical and cannot be described in terms of a classical treatment of the SPP and radiation fields~\cite{HOM}. Indeed, using a classical description of the fields leads to a minimum probability of 0.5~\cite{Ghosh}. An experimental test of this scattering scenario would yield strong evidence supporting the bosonic nature of SPPs in the frequency regime considered. We note that previous experiments~\cite{Alte,Lukin1,Kol} have tested some of the basic quantum properties of SPPs indirectly. The approach presented here would be more direct and conclusive.

In Fig.~\ref{fig9} we show the values of the coincidence probability obtained after the optimization procedure carried out according to Fig.~\ref{fig5}. One can clearly see how the coincidence drops to zero for values of the incidence angle approaching a 50:50 beamsplitter.

\section{Summary} 
In this work we introduced the quantum mechanical formalism for treating SPP interactions via scattering at an interface. To do this we extended a recent classical study to the quantum domain and at the same time generalized it to allow for the consideration of SPPs scattering at an angle. This work provides a first step in the direction of showing SPP quantum interference in addition to understanding SPP circuitry at the quantum level. Here, we used our theory to investigate the possibility of achieving plasmonic beamsplitters that operate faithfully at the single SPP-level. We found that a wide-range of splitting ratios can be reached depending on the frequency and angle of the input SPP excitations and that power loss due to unavoidable scattering into photon radiation can be suppressed to $5\%$ or even less in some cases. As the beamsplitter geometries we have investigated are readily accessible to experiments due to the basic properties of the materials involved, the theoretical findings reported here could be tested within the near-future. As an application of our theory, and an example of one such possible test, we investigated theoretically first-order quantum interference effects of SPPs using a 50:50 plasmonic beamsplitter optimized and incorporated into a Hong-Ou-Mandel type setup. Our study is the first to show that surface plasmon beamsplitters can reliably operate at the quantum level. Furthermore, our developed theory could be applied equally well to other waveguide geometries such as long-range, channel or dielectrically loaded~\cite{Hon,Ebbes,Boz}. Thus we expect our work to help open up new directions of research into the design of efficient and practical components for on-chip plasmonic-based QIP at the nanoscale.

\acknowledgments 
We thank A.\,V. Zayats, S. K. Ozdemir and C. Lee for discussions and comments on the manuscript. We acknowledge funding from ESF, EPSRC and QIPIRC.

\renewcommand{\theequation}{A-\arabic{equation}}
\setcounter{equation}{0}  
\section*{APPENDIX A}  
The positive and negative frequency parts of the electric and magnetic quantized radiation fields are given by
\bea
\hskip-0.08cm\hat{\bf E}_{r i}^+({{{\bf k}_i},q_i},{\bf r},t)&=& \left(\frac{\hbar c^2 q_i^2}{\epsilon_0 \pi \omega_i}\right)^{1/2}{\bm \phi}_r({\bf k}_i,q_i,z)\hat{A}_r({\bf k}_i,q_i)e^{-i \chi({\bf k}_i,q_i,{\bf r},t)} \nonumber \\
\hskip-0.08cm\hat{\bf E}_{r i}^-({{{\bf k}_i},q_i},{\bf r},t)&=& \left(\frac{\hbar c^2 q_i^2}{\epsilon_0 \pi \omega_i}\right)^{1/2}{\bm \phi}_r^{*}({\bf k}_i,q_i,z)\hat{A}_r^\dag({\bf k}_i,q_i)e^{i \chi({\bf k}_i,q_i,{\bf r},t)} \nonumber \\
\hskip-0.08cm\hat{\bf H}_{r i}^+({{{\bf k}_i},q_i},{\bf r},t)&=& \left(\frac{\epsilon_0\hbar c^6 q_i^2}{ \pi \omega_i^3}\right)^{1/2}{\bm \psi}_r({\bf k}_i,q_i,z)\hat{A}_r({\bf k}_i,q_i)e^{-i \chi({\bf k}_i,q_i,{\bf r},t)} \nonumber \\
\hskip-0.08cm\hat{\bf H}_{r i}^-({{{\bf k}_i},q_i},{\bf r},t)&=& \left(\frac{\epsilon_0 \hbar c^6 q_i^2}{ \pi \omega_i^3}\right)^{1/2}{\bm \psi}_r^{*}({\bf k}_i,q_i,z)\hat{A}_r^\dag({\bf k}_i,q_i)e^{i \chi({\bf k}_i,q_i,{\bf r},t)}\nonumber
\eea
with
\bea
&&\hskip-0.6cm{\bm \psi}_{{\textit {\tiny TM}}}({\bf k}_i,q_i,z)=\frac{k_i^2-\nu_i^2}{\nu_i}  \gamma_i^{\textit {\tiny TM}}   \bigg[e^{\nu_i z}\vartheta(-z)+  \nonumber \\
&&\hskip1.4cm (\cos q_i z + \eta_i^{-1}\sin q_i z)\vartheta(z) \bigg]\left(\frac{k_{y,i}}{k_i}\hat{\bf x}-\frac{k_{x,i}}{k_i}\hat{\bf y}\right) \nonumber
\eea
and 
\bea
&&\hskip-0.4cm{\bm \psi}_{{\textit {\tiny TE}}}({\bf k}_i,q_i,z)=i{\nu_i}  \gamma_i^{\textit {\tiny TE}}   \bigg[ (i \hat{{\bf k}}_i+\frac{k_i}{\nu_i}\hat{{\bf z}})e^{\nu_i z} \vartheta (-z)   \,+ \nonumber \\
&&\hskip-0.0cm\big[i \hat{{\bf k}}_i(\cos q_i z - \frac{q_i}{\nu_i} \sin q_i z)+\frac{k_i}{\nu_i}\hat{{\bf z}}(\cos q_i z + \frac{\nu_i}{q_i} \sin q_i z)\big] \vartheta (z) \bigg]. \nonumber
\eea
These can be found from the relation ${\bm \psi}_r({\bf k}_i,q_i,z)e^{-i \chi({\bf k}_i,q_i,{\bf r},t)}=\nabla \times {\bm \phi}_r({\bf k}_i,q_i,z)e^{i {\bf k}_i \cdot {\bf r}-i \omega_i t}$ and $\chi({\bf k}_i,q_i,{\bf r},t)= \omega_i t - {\bf k}_i \cdot {\bf r}-\pi/2$. For the SPP field we have 
\bea
\hat{\bf E}_{p i}^+({{{\bf k}_i}},{\bf r},t)&=& \left(\frac{\hbar \omega_i}{2\epsilon_0 p_i}\right)^{1/2}{\bm \phi}_p({\bf k}_i,z)\hat{a}({\bf k}_i)e^{-i \chi({\bf k}_i,{\bf r},t)}, \nonumber \\
\hat{\bf E}_{p i}^-({{{\bf k}_i}},{\bf r},t)&=& \left(\frac{\hbar \omega_i}{2\epsilon_0 p_i}\right)^{1/2}{\bm \phi}_p^{*}({\bf k}_i,z)\hat{a}^\dag({\bf k}_i)e^{i \chi({\bf k}_i,{\bf r},t)}, \nonumber\\
\hat{\bf H}_{p i}^+({{{\bf k}_i}},{\bf r},t)&=& \left(\frac{\epsilon_0 \hbar c^4}{2 \omega_i p_i}\right)^{1/2}{\bm \psi}_p({\bf k}_i,z)\hat{a}({\bf k}_i)e^{-i \chi({\bf k}_i,{\bf r},t)}, \nonumber \\
\hat{\bf H}_{p i}^-({{{\bf k}_i}},{\bf r},t)&=& \left(\frac{\epsilon_0 \hbar c^4}{2 \omega_i p_i}\right)^{1/2}{\bm \psi}_p^{*}({\bf k}_i,z)\hat{a}^\dag({\bf k}_i)e^{i \chi({\bf k}_i,{\bf r},t)},\nonumber 
\eea
where 
\bea
{\bm \psi}_p({\bf k}_i,z)&=&\frac{k_i^2-\nu_i^2}{\nu_i}\bigg[e^{\nu_i z}\vartheta(-z)+e^{-\nu_{0,i} z}\vartheta(z) \bigg] \left(\frac{k_{y,i}}{k_i}\hat{\bf x}-\frac{k_{x,i}}{k_i}\hat{\bf y}\right), \nonumber
\eea
which can be found from the relation ${\bm \psi}_p({\bf k}_i,z)e^{-i \chi({\bf k}_i,{\bf r},t)}=\nabla \times {\bm \phi}_p({\bf k}_i,z)e^{i {\bf k}_i\cdot{\bf r}-i \omega_i t}$ and $\chi({\bf k}_i,{\bf r},t)= \omega_i t - {\bf k}_i \cdot {\bf r}-\pi/2$.

The fields for an excitation defined by ${\bf k}_i$ and $q_i$ are given by $\hat{\bf E}_{\mu i}({{{\bf k}_i},q_i},{\bf r},t)=\hat{\bf E}_{\mu i}^-({{{\bf k}_i},q_i},{\bf r},t)+\hat{\bf E}_{\mu i}^+({{{\bf k}_i},q_i},{\bf r},t)$ and $\hat{\bf H}_{\mu i}({{{\bf k}_i},q_i},{\bf r},t)=\hat{\bf H}_{\mu i}^-({{{\bf k}_i},q_i},{\bf r},t)+\hat{\bf H}_{\mu i}^+({{{\bf k}_i},q_i},{\bf r},t)$. The total positive/negative fields are given by $\hat{\bf E}^{\pm}_{\mu i}({\bf r},t)=(2 \pi)^{-1}\iint_{0}^{q_{\rm cut}} {\rm d}{\bf k}_{i}{\rm d} q_i \hat{\bf E}^{\pm}_{\mu i}({{{\bf k}_i},q_i},{\bf r},t)$ and $\hat{\bf H}^{\pm}_{\mu i}({\bf r},t)=(2 \pi)^{-1}\iint_{0}^{q_{\rm cut}} {\rm d}{\bf k}_{i}{\rm d} q_i \hat{\bf H}^{\pm}_{\mu i}({{{\bf k}_i},q_i},{\bf r},t)$. The total fields are $\hat{\bf E}_{\mu i}({\bf r},t)=\hat{\bf E}^{+}_{\mu i}({\bf r},t)+\hat{\bf E}^{-}_{\mu i}({\bf r},t)$ and $\hat{\bf H}_{\mu i}({\bf r},t)=\hat{\bf H}^{+}_{\mu i}({\bf r},t)+\hat{\bf H}^{-}_{\mu i}({\bf r},t)$.

\renewcommand{\theequation}{B-\arabic{equation}}
\setcounter{equation}{0}  
\section*{APPENDIX B}  

For convenience we provide a compact analytical form for the coupling of Eq.~(\ref{couplcoeff}) given by
\bea
&&\hskip-0.4cm {\cal C}^{ji}_{\mu\nu}({\bf k}_j,q_j;{\bf k}_i,q_i)={\cal M}_{\mu,j } ({\bf k}_j,q_j){\cal M}^*_{\nu,i }({\bf k}_i,q_i)\times  \label{compactcouple} \\
&&\hskip2.5cm {\cal I}_{\mu \nu}({\bf k}_j,q_j;{\bf k}_i,q_i)\delta(k_{y,j}-k_{y,i})\delta(\omega_j-\omega_i),\nonumber
\eea
where 
\bea {\cal M}_{p,i }({\bf k}_i)&=&\sqrt{\frac{ \epsilon_0\omega_i  \epsilon_{d,i}\epsilon_{m,i}^2 \nu_{0,i}}{\pi k_{x,i}^* (\epsilon_{m,i}^2-\epsilon_{d,i}^2) }   \frac{k_i^*}{k_i} }, \nonumber \\ {\cal M}_{{\textit {\tiny TM}} ,i}({\bf k}_i,q_i)&=&  \sqrt{\frac{ \epsilon_0 \omega_i  \epsilon_{d,i}}{4 \pi^2 k_{x,i}^*}    \frac{k_i^*}{k_i}}, \nonumber \\  {\cal M}_{{\textit {\tiny TE}},i }({\bf k}_i,q_i)  &=& \sqrt{  \frac{1}{4 \pi^2  \epsilon_0  \omega_i  k_{x,i}}   \frac{k_i}{k_i^*}} \nonumber  \eea
and similarly for the $j$ terms, with 
\bea
&&\hskip-0.4cm{\cal I}_{p\,p}({\bf k}_j;{\bf k}_i)=\frac{2\pi k_{x,i}^*}{\epsilon_0 \omega}   \frac{k_j}{k_i^*}\left[ \frac{1}{\epsilon_{m,j}(\nu_j+\nu_i)}+\frac{1}{\epsilon_{d,j}(\nu_{0,j}+\nu_{0,i})} \right] \nonumber \\
&&\hskip-0.4cm{\cal I}_{{\textit {\tiny TM}} \,p }({\bf k}_j,q_j;{\bf k}_i)=\frac{2\pi k_{x,i}^*}{\epsilon_0 \omega}   \frac{k_j}{k_i^*} \times   \nonumber \\
&&\hskip-0.0cm \bigg[\epsilon_{d,j}^{-1}\big[\, \frac{1}{\nu_{0,i} + i q_j } - \frac{ r_{{\textit {\tiny TM}} , j}(k_j,q_j) }{ \nu_{0,i} - i q_j }\big]    +   \epsilon_{m,j}^{-1}  \frac{ (1-r_{{\textit {\tiny TM}} , j}(k_j,q_j)) }{\nu_i   + \nu_j   } \bigg],   \nonumber \\
&&\hskip-0.4cm{\cal I}_{p   \, {\textit {\tiny TM}} }({\bf k}_j;{\bf k}_i,q_i)=\frac{2\pi k_{x,i}^*}{\epsilon_0 \omega}   \frac{k_j}{k_i^*} \times \nonumber \\
&&\hskip-0.0cm \bigg[\epsilon_{d,j}^{-1}\big[\,  \frac{1}{\nu_{0,j} - i q_i } -\frac{ r_{{\textit {\tiny TM}} , i}^*(k_i,q_i) }{ \nu_{0,j} + i q_i }\big]    +   \epsilon_{m,j}^{-1}  \frac{ (1-r_{{\textit {\tiny TM}} , i}^*(k_i,q_i)) }{\nu_j   + \nu_i   } \bigg],      \nonumber       \\
&&\hskip-0.4cm{\cal I}_{{\textit {\tiny TM}} \, {\textit {\tiny TM}} }({\bf k}_j,q_j;{\bf k}_i,q_i)=\frac{2\pi k_{x,i}^*}{\epsilon_0 \omega}   \frac{k_j}{k_i^*} \times \nonumber \\
&&\hskip-0.0cm \bigg[\epsilon_{d,j}^{-1}   \, 2\pi \,   [\, r_{{\textit {\tiny TM}} , j}(k_j,q_j) r_{{\textit {\tiny TM}} , i}^*(k_i,q_i)    \,   \delta_+(q_j-q_i)   \nonumber  \\
&&\hskip0.2cm  - r_{ {\textit {\tiny TM}} , j} (k_j,q_j)  \,  \delta_+(q_j + q_i) - r_{{\textit {\tiny TM}} , i}^*(k_i,q_i)   \, \delta_+(-q_j - q_i)    \nonumber \\
&&\hskip2.2cm+    \delta_+(-q_j+q_i)]  +\epsilon_{m,j}^{-1}  (1-r_{{\textit {\tiny TM}} , j}(k_j,q_j))  \times \nonumber\\
&&\hskip4.2cm(1-r_{{\textit {\tiny TM}} , i}^*(k_i,q_i)) \frac{1}{\nu_j   + \nu_i   } \bigg],   \nonumber \\
&&\hskip-0.4cm{\cal I}_{{\textit {\tiny TE}} \,p }({\bf k}_j,q_j;{\bf k}_i)=0,      \nonumber       \\
&&\hskip-0.4cm{\cal I}_{p   \, {\textit {\tiny TE}} }({\bf k}_j;{\bf k}_i,q_i)=\frac{2\pi}{\epsilon_0 \omega}    \bigg[ \bigg(\frac{k_{y,j} \, i \nu_{0,j} k_i^*}{\epsilon_{d,j} k_j } -  \frac{k_{y,i}  q_i k_j}{\epsilon_{d,j} k_i ^*}   \bigg) \frac{ r_{{\textit {\tiny TE}} , i}^*(k_i,q_i) }{ \nu_{0,j} + i q_i }      \nonumber \\
&&\hskip0.3cm - \bigg(\frac{k_{y,j} \, i \nu_{0,j} k_i^*}{\epsilon_{d,j} k_j } +  \frac{k_{y,i} q_i k_j}{\epsilon_{d,j} k_i ^*}   \bigg) \frac{1 }{ \nu_{0,j} - i q_i }      \nonumber \\
&&\hskip0.3cm + \bigg(\frac{k_{y,j} \, i \nu_{j} k_i^*}{\epsilon_{m,j} k_j } -  \frac{k_{y,i} \, (i \nu_{ i})^* k_j}{\epsilon_{m,j} k_i ^*}   \bigg) \frac{(1-r_{{\textit {\tiny TE}} , i}^*(k_i,q_i)) }{\nu_j   + \nu_i   } \bigg],      \nonumber       \\
&&\hskip-0.4cm{\cal I}_{{\textit {\tiny TE}} \, {\textit {\tiny TE}} }({\bf k}_j,q_j;{\bf k}_i,q_i)= 2\pi k_{x,j} \epsilon_0 \omega   \frac{k_i^*}{k_j} \times \nonumber \\
&&\hskip0.0cm \bigg[  r_{{\textit {\tiny TE}} , j}(k_j,q_j) r_{{\textit {\tiny TE}} , i}^*(k_i,q_i) \,  2\pi \,   \delta_+(q_j-q_i)   \nonumber  \\
&&\hskip0.0cm  - r_{ {\textit {\tiny TE}} , j} (k_j,q_j)  \, 2\pi \,   \delta_+(q_j + q_i) - r_{{\textit {\tiny TE}} , i}^*(k_i,q_i) \, 2\pi \,  \delta_+(-q_j - q_i)    \nonumber \\
&&\hskip2.2cm+ 2\pi \, \delta_+(-q_j+q_i)   +   (1-r_{{\textit {\tiny TE}} , j}(k_j,q_j))  \times \nonumber\\
&&\hskip4.2cm(1-r_{{\textit {\tiny TE}} , i}^*(k_i,q_i)) \frac{1}{\nu_j   + \nu_i   } \bigg],   \nonumber \\
&&\hskip-0.4cm{\cal I}_{{\textit {\tiny TE}} \,{\textit {\tiny TM}} }({\bf k}_j,q_j;{\bf k}_i, q_i)=0,     \nonumber  \\
&&\hskip-0.4cm{\cal I}_{{\textit {\tiny TM}} \, {\textit {\tiny TE}} }({\bf k}_j,q_j;{\bf k}_i,q_i)=\frac{2\pi}{\epsilon_0 \omega}  \bigg[  \bigg( - \frac{k_{y,j} \, q_{j} k_i^*}{\epsilon_{d,j} k_j }  +   \frac{k_{y,i} \,q_{i} k_j}{\epsilon_{d,j} k_i ^*}  \bigg)  \times \nonumber \\
&&\hskip1cm  r_{{\textit {\tiny TM}} , j}(k_j,q_j) r_{{\textit {\tiny TE}} , i}^*(k_i,q_i) \, 2\pi \, \delta_+(q_j - q_i )      \nonumber \\
&&\hskip0.3cm + \bigg(  \frac{k_{y,j} \, q_{j} k_i^*}{\epsilon_{d,j} k_j } +  \frac{k_{y,i} \,q_{i} k_j}{\epsilon_{d,j} k_i^* }   \bigg)   r_{{\textit {\tiny TM}} , j}(k_j,q_j) \, 2\pi \,  \delta_+(q_j + q_i )      \nonumber \\
&&\hskip0.3cm - \bigg(  \frac{k_{y,j} \, q_{j} k_i^*}{\epsilon_{d,j} k_j } +  \frac{k_{y,i} \,q_{i} k_j}{\epsilon_{d,j} k_i ^* }   \bigg)  r_{{\textit {\tiny TE}} , i}^*(k_i,q_i) \, 2\pi \, \delta_+(-q_j - q_i )      \nonumber \\
&&\hskip0.3cm + \bigg(  \frac{k_{y,j} \, q_{j} k_i^*}{\epsilon_{d,j} k_j } -  \frac{k_{y,i} \,q_{i} k_j}{\epsilon_{d,j} k_i^* }   \bigg) \, 2\pi \,    \delta_+(-q_j + q_i )      \nonumber \\
&&\hskip0.3cm +  \bigg(\frac{k_{y,j} \, i \nu_{j} k_i^*}{\epsilon_{m,j} k_j } - \frac{k_{y,i} \, (i \nu_{ i})^* k_j}{\epsilon_{m,j} k_i^* }   \bigg) \times \nonumber \\
&&\hskip1cm (1-r_{{\textit {\tiny TM}} , j}(k_j,q_i)) (1-r_{{\textit {\tiny TE}} , i}^*(k_i,q_i))  \frac{1}{\nu_j   + \nu_i   } \bigg].      \nonumber  
\eea 
Here we have $r_{{\textit {\tiny TM}}\,i}(k_i,q_i)=(i\epsilon_{d,i}\nu_i-\epsilon_{m,i}q_i)/(i\epsilon_{d,i}\nu_i+\epsilon_{m,i}q_i)$, as well as $r_{{\textit {\tiny TE}}\,i}(k_i,q_i)=(i\nu_i-q_i)/(i\nu_i+q_i)$, with similar definitions for side $j$ together with the function
$\delta_+(k) = \frac{1}{2\pi} \int_{-\infty}^{\infty}h(z)e^{i k z}dz= \frac{1}{2} \delta(k) - \frac{1}{2\pi i} \frac{\cal P}{k}$,
where $h(z)$ represents the Heaviside function, $\cal P$ stands for the Cauchy principal value of the integral and $\delta(k) = \frac{1}{2\pi} \int_{-\infty}^{\infty} e^{i k z}dz$. 

Note that each of the above couplings will be integrated over $k_x$, $k_y$ and $q$ for the relevant region due to the integrals appearing in Eqs.~(\ref{electrict})~and~(\ref{magnetict}). This gives physical meaning to the delta functions which select out the operators specified in the coupled operator equations, Eqs.~(\ref{couple1}) and (\ref{couple2TM})-(\ref{couple4TE}). Here, the integration over $k_x$ is changed to integration over $\omega$ for convenience. For example, for SPPs we have $ {\rm d} k_{x,i} \to (k_i/k_{x,i}V_{G,i}){\rm d} \omega_i$ with the operators transforming accordingly, $\hat{a}^f({\bf k}_i) \to (k_{x,i}V_{G,i}/k_i)^{1/2}\hat{a}^f(\omega_i,k_{y,i})$. Thus, all operators are strictly transformed from wavenumber to frequency domain for the $k_x$ component of the wavevector, hence the notation $\hat{a}^f({\bm \omega_i})$ for the operators, where ${\bm \omega_i}=(\omega_i,k_{y,i})$. 

\renewcommand{\theequation}{C-\arabic{equation}}
\setcounter{equation}{0}  
\section*{APPENDIX C}  

The $D$ and $F$ matrices appearing in the set of coupled matrix equations (\ref{couple5TM})-(\ref{couple6TE}) are defined in terms of the elements of $C$ matrices, which are themselves defined in terms of the couplings $ {\cal C}^{ji}_{\mu\nu}({\bf k}_j,q_j;{\bf k}_i,q_i)$ (specified by Eq.~(\ref{couplcoeff}) and given in Appendix B) as follows:
\bea
&& (C_{{\textit {\tiny TM}}\, {\textit {\tiny TM}}})_{00}={\cal C}^{ji}_{pp}({\bf k}_j;{\bf k}_i), \label{coupleelem} \\
&& (C_{{\textit {\tiny TM}}\, {\textit {\tiny TM}}})_{0n}={\cal C}^{ji}_{p\, {\textit {\tiny TM}}}({\bf k}_j;{\bf k}_i,u_n)\sqrt{w_n'}, \nonumber \\ 
&& (C_{{\textit {\tiny TM}}\, {\textit {\tiny TM}}})_{m0}={\cal C}^{ji}_{{\textit {\tiny TM}} \, p}({\bf k}_j,u_m;{\bf k}_i)\sqrt{w_m'}, \nonumber \\
&& (C_{{\textit {\tiny TM}}\, {\textit {\tiny TE}}})_{0n}={\cal C}^{ji}_{p\, {\textit {\tiny TE}}}({\bf k}_j;{\bf k}_i,u_n)\sqrt{w_n'}, \nonumber \\ 
&& (C_{rr})_{nn}={\cal C}^{ji}_{rr}({\bf k}_j,u_n;{\bf k}_i,u_n), \nonumber \\ 
&& (C_{rr})_{mn}={\cal C}^{ji}_{rr}({\bf k}_j,u_m;{\bf k}_i,u_n)\sqrt{w_m'w_n'}, \quad m\neq n, \nonumber
\eea
with zero elements 
\bea
&& (C_{{\textit {\tiny TE}}\, {\textit {\tiny TM}}})_{mn}=0~\forall m,n, \label{coupleelem2} \\
&&(C_{{\textit {\tiny TM}}\, {\textit {\tiny TE}}})_{00}=(C_{{\textit {\tiny TM}}\, {\textit {\tiny TE}}})_{m0}=0, \nonumber \\
&&(C_{{\textit {\tiny TE}}\, {\textit {\tiny TE}}})_{00}= (C_{{\textit {\tiny TE}}\, {\textit {\tiny TE}}})_{0n}=(C_{{\textit {\tiny TE}}\, {\textit {\tiny TE}}})_{m0}=0. \nonumber
\eea
From the above matrices we then have the $D$ matrices
\bea
&&(D_{{\textit {\tiny TM}}\, {\textit {\tiny TM}}} ) _{00} = (C_{{\textit {\tiny TM}}\, {\textit {\tiny TM}}} ) _{00} \label{DTMTM} \\
&&(D_{{\textit {\tiny TM}}\, {\textit {\tiny TM}}} ) _{0n} = \left\{
\begin{array}{rcc} 
(C_{{\textit {\tiny TM}}\, {\textit {\tiny TM}}} ) _{0n} & n\leq mmax_i & \\
i \,(C_{{\textit {\tiny TM}}\, {\textit {\tiny TM}}} ) _{0n} & n> mmax_i & k_i\in \mathbb{R}\\
-i \,(C_{{\textit {\tiny TM}}\, {\textit {\tiny TM}}} ) _{0n} & n> mmax_i & k_i\notin \mathbb{R}\\
\end{array} \right. \nonumber \\
&&(D_{{\textit {\tiny TM}}\, {\textit {\tiny TM}}} ) _{m0} = (C_{{\textit {\tiny TM}}\, {\textit {\tiny TM}}} ) _{m0} \nonumber \\
&&(D_{{\textit {\tiny TM}}\, {\textit {\tiny TM}}} ) _{mn} = \left\{
\begin{array}{rcc} 
(C_{{\textit {\tiny TM}}\, {\textit {\tiny TM}}} ) _{mn} & n\leq mmax_i & \\
i \,(C_{{\textit {\tiny TM}}\, {\textit {\tiny TM}}} ) _{mn} & n> mmax_i & k_i\in \mathbb{R}\\
-i \,(C_{{\textit {\tiny TM}}\, {\textit {\tiny TM}}} ) _{mn} & n> mmax_i & k_i\notin \mathbb{R}\\
\end{array} \right. \nonumber \\
&&(D_{r\, {\textit {\tiny TE}}} ) _{00} = (C_{r\, {\textit {\tiny TE}}} ) _{00} \nonumber \\
&&(D_{r\, {\textit {\tiny TE}}} ) _{0n} = \left\{
\begin{array}{rcc} 
(C_{r\, {\textit {\tiny TE}}} ) _{0n} & n\leq mmax_i & \\
-i \,(C_{r\, {\textit {\tiny TE}}} ) _{0n} & n> mmax_i & k_i\in \mathbb{R}\\
i \,(C_{r\, {\textit {\tiny TE}}} ) _{0n} & n> mmax_i & k_i\notin \mathbb{R}\\
\end{array} \right. \nonumber \\
&&(D_{r\, {\textit {\tiny TE}}} ) _{m0} = (C_{r\, {\textit {\tiny TE}}} ) _{m0} \nonumber \\
&&(D_{r\, {\textit {\tiny TE}}} ) _{mn} = \left\{
\begin{array}{rcc} 
(C_{r\, {\textit {\tiny TE}}} ) _{mn} & n\leq mmax_i & \\
-i \,(C_{r\, {\textit {\tiny TE}}} ) _{mn} & n> mmax_i & k_i\in \mathbb{R}\\
i \,(C_{r\, {\textit {\tiny TE}}} ) _{mn} & n> mmax_i & k_i\notin \mathbb{R}\\
\end{array} \right. \nonumber \\
&& (D_{{\textit {\tiny TE}}\, {\textit {\tiny TM}}})_{mn}=0~ \quad \forall m,n, \nonumber 
\eea
and the $F$ matrices
\bea
&&(F_{{\textit {\tiny TM}}\, r} ) _{00} = (C_{{\textit {\tiny TM}}\, r} ) _{00} \label{FTMr} \\
&&(F_{{\textit {\tiny TM}}\,r} ) _{0n} = (C_{{\textit {\tiny TM}}\,r} ) _{0n} \nonumber \\
&&(F_{{\textit {\tiny TM}}\, r} ) _{m0} = \left\{
\begin{array}{rcc} 
(C_{{\textit {\tiny TM}}\, r} ) _{m0} & m\leq mmax_j & \\
i \,(C_{{\textit {\tiny TM}}\, r} ) _{m0} & m> mmax_j & k_j\in \mathbb{R}\\
-i \,(C_{{\textit {\tiny TM}}\, r} ) _{m0} & m> mmax_j & k_j\notin \mathbb{R}\\
\end{array} \right. \nonumber \\
&&(F_{{\textit {\tiny TM}}\, r} ) _{mn} = \left\{
\begin{array}{rcc} 
(C_{{\textit {\tiny TM}}\, r} ) _{mn} & m\leq mmax_j & \\
i \,(C_{{\textit {\tiny TM}}\, r} ) _{mn} & m> mmax_j & k_j\in \mathbb{R}\\
-i \,(C_{{\textit {\tiny TM}}\, r} ) _{mn} & m> mmax_j & k_j\notin \mathbb{R}\\
\end{array} \right. \nonumber  \\
&&(F_{{\textit {\tiny TE}}\,  {\textit {\tiny TE}}} ) _{00} = (C_{{\textit {\tiny TE}}\,  {\textit {\tiny TE}}} ) _{00} \nonumber \\
&&(F_{{\textit {\tiny TE}}\,  {\textit {\tiny TE}}} ) _{0n} = (C_{{\textit {\tiny TE}}\,  {\textit {\tiny TE}}} ) _{0n} \nonumber \\
&&(F_{{\textit {\tiny TE}}\,  {\textit {\tiny TE}}} ) _{m0} = \left\{
\begin{array}{rcc} 
(C_{{\textit {\tiny TE}}\,  {\textit {\tiny TE}}} ) _{m0} & m\leq mmax_j & \\
-i \,(C_{{\textit {\tiny TE}}\,  {\textit {\tiny TE}}} ) _{m0} & m> mmax_j & k_j\in \mathbb{R}\\
i \,(C_{{\textit {\tiny TE}}\,  {\textit {\tiny TE}}} ) _{m0} & m> mmax_j & k_j\notin \mathbb{R}\\
\end{array} \right. \nonumber \\
&&(F_{{\textit {\tiny TE}}\,  {\textit {\tiny TE}}} ) _{mn} = \left\{
\begin{array}{rcc} 
(C_{{\textit {\tiny TE}}\,  {\textit {\tiny TE}}} ) _{mn} & m\leq mmax_j & \\
-i \,(C_{{\textit {\tiny TE}}\,  {\textit {\tiny TE}}} ) _{mn} & m> mmax_j & k_j\in \mathbb{R}\\
i \,(C_{{\textit {\tiny TE}}\,  {\textit {\tiny TE}}} ) _{mn} & m> mmax_j & k_j\notin \mathbb{R}\\
\end{array} \right. \nonumber  \\
&& (F_{{\textit {\tiny TE}}\, {\textit {\tiny TM}}})_{mn}=0~ \quad \forall m,n. \label{FTETM} \nonumber
\eea

Here $mmax_i$ ($mmax_j$) represents the number of non-evanescent (propagating) radiation excitations in region $i$ ($j$) upon discretization. The imaginary factors account for the different normalization of non-evanescent and evanescent excitations, as described in the main text. Note that the subscripts $TM$ and $TE$ denote a different $D$ or $F$ matrix, whereas the indices $m$ and $n$ denote the elements of these matrices. In the limit of normal incidence, {\it i.e.} $\theta_{i_i}=0$, coupling to TE excitations does not occur and the $D$ and $F$ matrices lead to a transfer matrix as given previously in Ref.~\cite{Oult}. 

\renewcommand{\theequation}{D-\arabic{equation}}
\setcounter{equation}{0}  
\section*{APPENDIX D}  

The $T_{ij}$ entries of the transfer matrix are given by
\bea
T_{31}&=&\chi[D_{{\textit {\tiny TM}}\, {\textit {\tiny TE}}}^TF_{{\textit {\tiny TM}}\, {\textit {\tiny TM}}}^*(\openone+D_{{\textit {\tiny TM}}\, {\textit {\tiny TM}}}^TF_{{\textit {\tiny TM}}\, {\textit {\tiny TM}}}^*)^{-1} \times \nonumber \\
&&\hskip0.5cm(D_{{\textit {\tiny TM}}\, {\textit {\tiny TM}}}^TF_{{\textit {\tiny TM}}\, {\textit {\tiny TM}}}^*-\openone)-D_{{\textit {\tiny TM}}\, {\textit {\tiny TE}}}^TF_{{\textit {\tiny TM}}\, {\textit {\tiny TM}}}^*], \nonumber \\
T_{32}&=&\chi[2D_{{\textit {\tiny TM}}\, {\textit {\tiny TE}}}^TF_{{\textit {\tiny TM}}\, {\textit {\tiny TM}}}^*(\openone+D_{{\textit {\tiny TM}}\, {\textit {\tiny TM}}}^TF_{{\textit {\tiny TM}}\, {\textit {\tiny TM}}}^*)^{-1}D_{{\textit {\tiny TM}}\, {\textit {\tiny TM}}}^T -2D_{{\textit {\tiny TM}}\, {\textit {\tiny TE}}}^T],\nonumber \\
T_{33}&=&\chi[\openone-D_{{\textit {\tiny TM}}\, {\textit {\tiny TE}}}^TF_{{\textit {\tiny TM}}\, {\textit {\tiny TE}}}^*-D_{{\textit {\tiny TE}}\, {\textit {\tiny TE}}}^TF_{{\textit {\tiny TE}}\, {\textit {\tiny TE}}}^* \nonumber \\
&&\hskip0.5cm+D_{{\textit {\tiny TM}}\, {\textit {\tiny TE}}}^TF_{{\textit {\tiny TM}}\, {\textit {\tiny TM}}}^*(\openone+D_{{\textit {\tiny TM}}\, {\textit {\tiny TM}}}^TF_{{\textit {\tiny TM}}\, {\textit {\tiny TM}}}^*)^{-1}D_{{\textit {\tiny TM}}\, {\textit {\tiny TM}}}^TF_{{\textit {\tiny TM}}\, {\textit {\tiny TE}}}^*], \nonumber \\
T_{34}&=&2\chi D_{{\textit {\tiny TE}}\, {\textit {\tiny TE}}}^T, \nonumber 
\eea
where $\chi=[\openone+D_{{\textit {\tiny TM}}\, {\textit {\tiny TE}}}^TF_{{\textit {\tiny TM}}\, {\textit {\tiny TE}}}^*+D_{{\textit {\tiny TE}}\, {\textit {\tiny TE}}}^TF_{{\textit {\tiny TE}}\, {\textit {\tiny TE}}}^*-D_{{\textit {\tiny TM}}\, {\textit {\tiny TE}}}^TF_{{\textit {\tiny TM}}\, {\textit {\tiny TM}}}^*(\openone+D_{{\textit {\tiny TM}}\, {\textit {\tiny TM}}}^TF_{{\textit {\tiny TM}}\, {\textit {\tiny TM}}}^*)^{-1}D_{{\textit {\tiny TM}}\, {\textit {\tiny TM}}}^TF_{{\textit {\tiny TM}}\, {\textit {\tiny TE}}}^*]^{-1}$. Then we have $T_{41}=F_{{\textit {\tiny TE}}\, {\textit {\tiny TE}}}^*T_{31},~T_{42}=F_{{\textit {\tiny TE}}\, {\textit {\tiny TE}}}^*T_{32},~T_{43}=F_{{\textit {\tiny TE}}\, {\textit {\tiny TE}}}^*(\openone+T_{33}),~T_{44}=(F_{{\textit {\tiny TE}}\, {\textit {\tiny TE}}}^*T_{34}-\openone)$ together with
\bea
T_{11}&=&(\openone+D_{{\textit {\tiny TM}}\, {\textit {\tiny TM}}}^TF_{{\textit {\tiny TM}}\, {\textit {\tiny TM}}}^*)^{-1}(D_{{\textit {\tiny TM}}\, {\textit {\tiny TM}}}^TF_{{\textit {\tiny TM}}\, {\textit {\tiny TM}}}^* -\openone+D_{{\textit {\tiny TM}}\, {\textit {\tiny TM}}}^TF_{{\textit {\tiny TM}}\, {\textit {\tiny TE}}}^*T_{31}), \nonumber \\
T_{12}&=&(\openone+D_{{\textit {\tiny TM}}\, {\textit {\tiny TM}}}^TF_{{\textit {\tiny TM}}\, {\textit {\tiny TM}}}^*)^{-1}(2D_{{\textit {\tiny TM}}\, {\textit {\tiny TM}}}^T+D_{{\textit {\tiny TM}}\, {\textit {\tiny TM}}}^TF_{{\textit {\tiny TM}}\, {\textit {\tiny TE}}}^*T_{32}), \nonumber \\
T_{13}&=&(\openone+D_{{\textit {\tiny TM}}\, {\textit {\tiny TM}}}^TF_{{\textit {\tiny TM}}\, {\textit {\tiny TM}}}^*)^{-1}D_{{\textit {\tiny TM}}\, {\textit {\tiny TM}}}^TF_{{\textit {\tiny TM}}\, {\textit {\tiny TE}}}^*(\openone+T_{33}), \nonumber \\
T_{14}&=&(\openone+D_{{\textit {\tiny TM}}\, {\textit {\tiny TM}}}^TF_{{\textit {\tiny TM}}\, {\textit {\tiny TM}}}^*)^{-1}D_{{\textit {\tiny TM}}\, {\textit {\tiny TM}}}^TF_{{\textit {\tiny TM}}\, {\textit {\tiny TE}}}^*T_{34}, \nonumber \\
T_{21}&=&F_{{\textit {\tiny TM}}\, {\textit {\tiny TM}}}^*(\openone-T_{11})+F_{{\textit {\tiny TM}}\, {\textit {\tiny TE}}}^*T_{31}, \nonumber \\
T_{22}&=&\openone+F_{{\textit {\tiny TM}}\, {\textit {\tiny TE}}}^*T_{32}-F_{{\textit {\tiny TM}}\, {\textit {\tiny TM}}}^*T_{12}, \nonumber \\
T_{23}&=&F_{{\textit {\tiny TM}}\, {\textit {\tiny TE}}}^*(\openone+T_{32})-F_{{\textit {\tiny TM}}\, {\textit {\tiny TM}}}^*T_{13}, \nonumber \\
T_{24}&=&F_{{\textit {\tiny TM}}\, {\textit {\tiny TE}}}^*T_{34}-F_{{\textit {\tiny TM}}\, {\textit {\tiny TM}}}^*T_{14}. \nonumber 
\eea



\begin{thebibliography}{99}

\bibitem{Zayats} A. V. Zayats, I. I. Smolyaninov and A. A. Maradudin, Phys. Rep. {\bf 408}, 131 (2005); S. A. Maier, Plasmonics: Fundamentals and Applications, (Springer, New York, 2007).

\bibitem{photoncircuit} W. L. Barnes, A. Dereux and T. W. Ebbesen, Nature {\bf 424}, 824 (2003).

\bibitem{plasmonQIP} J. L. van Velsen, J. Tworzydlo and C. W. J. Beenakker, Phys. Rev. A {\bf 68} 043807 (2003); S. Fasel, M. Halder, N. Gisin and H. Zbinden, New J. Phys. {\bf 8}, 13 (2006); X.-F. Ren, G.-P. Guo, Y.-F. Huang, Z.-W. Wang, G.-C. Guo, Opt. Lett. {\bf 31}, 2792 (2006); A. Kamli, S. A. Moiseev and B. C. Sanders, Phys. Rev. Lett. {\bf 101}, 263601 (2008). 

\bibitem{Alte} E. Altewischer, M. P. van Exter and J. P. Woerdman, Nature {\bf 418}, 304 (2002); E. Moreno, F. J. Garc'a-Vidal, D. Erni, J. I. Cirac and L. Mart'n-Moreno, Phys. Rev. Lett. {\bf 92}, 236801 (2004); S. Fasel, F. Robin, E. Moreno, D. Erni, N. Gisin and H. Zbinden, Phys. Rev. Lett. {\bf 94}, 110501 (2005); X.-F. Ren, G.-P. Guo, Y.-F. Huang, C.-F. Li and G.-C. Guo, Europhys. Lett. {\bf 76}, 753 (2006); A. Huck, S. Smolka, P. Lodahl, A. S. Sorensen, A. Boltasseva, J. Janousek and U. L. Andersen, Phys. Rev. Lett. {\bf 102}, 246802 (2009).

\bibitem{Lukin1} D. E. Chang, A. S. Sorensen, P. R. Hemmer and M. D. Lukin, Phys. Rev. Lett. {\bf 97}, 053002 (2006); A. V. Akimov, A. Mukherjee, C. L. Yu, D. E. Chang, A. S. Zibrov, P. R. Hemmer, H. Park, M. D. Lukin, Nature {\bf 450}, 402 (2007).

\bibitem{Lukin2} D. E. Chang, A. S. Sorensen, E. A. Demler, M. D. Lukin, Nature Phys. {\bf 3}, 807 (2007).

\bibitem{TSPP} M. S. Tame, C. Lee, J. Lee, D. Ballester, M. Paternostro, A. V. Zayats and M. S. Kim, Phys. Rev. Lett. {\bf 101}, 190504 (2008).

\bibitem{DSPP} D. Ballester, M. S. Tame, C. Lee, J. Lee and M. S. Kim, Phys. Rev. A {\bf 79}, 053845 (2009).

\bibitem{Lukin3} A. L. Falk, F. H. L. Koppens, C. L. Yu, K. Kang, N. de Leon Snapp, A. V. Akimov, M.-H. Jo, M. D. Lukin, H. Park, Nature Phys. {\bf 5}, 475 (2009).

\bibitem{Kol} R. Kolesov, B. Grotz, G. Balasubramanian, R. J. St\"ohr, A. A. L. Nicolet, P. R. Hemmer, F. Jelezko and J. Wrachtrup, Nature Phys. {\bf 5}, 470 (2009).

\bibitem {Gramotnev} D. K. Gramotnev and S. I. Bozhevolnyi, Nature Photonics {\bf 4}, 83 (2010).

\bibitem{Maier2} S. A. Maier, IEEE J. Sel. Top. Quant. Elec. {\bf 12}, 1214 (2006); S. A. Maier ibid. {\bf 12}, 1671 (2006).

\bibitem{Oult} R. F. Oulton, D. F. Pile, Y. Liu, and X. Zhang, Phys. Rev. B {\bf 76}, 035408 (2007).

\bibitem{Elser} J. Elser and V. A. Podolskiy, Phys. Rev. Lett. {\bf 100}, 066402 (2008); M. Zhong-Tuan, W. Pei, C. Yong, T. Hong-Gao and M. Hai, Chin. Phys. Lett. {\bf 23}, 2545 (2006); Q. Bai, J. Chen, C. Liu, J. Xu, C. Cheng, N.-H. Shen and H.-T. Wang, Phys. Rev. B {\bf 79}, 155401 (2009); T. V\'{a}ry and P. Marko\u{s}, Physica B {\bf 405}, 2982 (2010); T. V\'{a}ry and P. Marko\u{s}, Proc.  SPIE {\bf 7353}, 73530K (2009).

\bibitem{Ditl} H. Ditlbacher, J. R. Krenn, G. Schider, A. Leitner and F. R. Aussenegg, App. Phys. Lett. {\bf 81}, 1762 (2002).

\bibitem{Hon} A. Hohenau, J. R. Krenn, A. L. Stepanov, A. Drezet, H. Ditlbacher, B. Steinberger, A. Leitner and F. R. Aussenegg, Opt. Lett. {\bf 30} (8), 893 (2003).

\bibitem{Ebbes} T. W. Ebbesen, C. Genet and S. I. Bozhevolnyi, Physics Today, {\bf 44} (May 2008); S. Bozhevolnyi, Ed., Plasmonic waveguides (World Scientific, Singapore, 2008).

\bibitem{Boz} S. I. Bozhevolnyi, V. S. Volkov, E. Devaux, J.-Y. Laluet, T. W. Ebbesen, Nature {\bf 440}, 508 (2006).

\bibitem{Steg} G. I. Stegeman, A. A. Maradudin and T. S. Rahman, Phys. Rev. B {\bf 23}, 2576 (1981); G. I. Stegeman, N. E. Glass, A. A. Maradudin, T. P. Shen and R. F. Wallis, Opt. Lett. {\bf 8}, 626 (1983); G. I. Stegeman, A. A. Maradudin, T. P. Shen and R. F. Wallis, Phys. Rev. B {\bf 29}, 6530 (1984).

\bibitem{Loudon} R. Loudon, {\sl The Quantum Theory of Light}, 3$^{\rm rd}$ Ed., Oxford University Press, Oxford (2000). 

\bibitem{ER} J.\,M.\,Elson\,and\,R.\,H.\,Ritchie,\,Phys.\,Rev.\,B\,{\bf 4},\,4129\,(1971).

\bibitem{SnyderLove} A. W. Snyder and J. D. Love, {\sl Optical Waveguide Theory}, 1$^{\rm st}$ Ed., Chapman and Hall, New York (1983).

\bibitem{Oultrec} To recover the form given in Ref.~\cite{Oult}: $q_i \epsilon_{m,i} / \nu_i \epsilon_{d,i}= i(r_i({\bf k}_i)-1)/(r_i({\bf k}_i)+1)$, with  $r_i({\bf k}_i)=(\epsilon_{d,i}k_{zm,i}+\epsilon_{m,i}k_{zd,i})/(\epsilon_{d,i}k_{zm,i}-\epsilon_{m,i}k_{zd,i})$, $k_{zd,i}=q_i$ and $k_{zm,i}=i\nu_i$.

\bibitem{SPPQ} J. Nkoma, R. Loudon and D. R. Tilley, J. Phys. C: Solid State Phys. {\bf 7}, 3547 (1974); M. S. Toma$\check{\rm s}$ and M. $\check{\rm S}$unji$\acute{\rm c}$, Phys. Rev. B {\bf 12} 5363 (1975); Y. O. Nakamura, Prog. Theor. Phys. {\bf 70}, 908 (1983).

\bibitem{JohnChrist} P. B. Johnson and R. W. Christy, Phys. Rev. B {\bf 6}, 4370 (1972).

\bibitem{Schev} V. V. Schevchenko, {\sl Continuous Transitions in Open Waveguides}, The Golem Press, Boulder, CO (1971).

\bibitem{Land} L. D. Landau and E. M. Lifshitz, {\sl Electrodynamics of continuous media}, 2$^{\rm nd}$ Ed., Pergamon Press, Oxford (1984).

\bibitem{Huttner} B. Huttner and S. M. Barnett, Phys. Rev. A {\bf 46}, 4306 (1992).

\bibitem{Loudon2} M. Artoni and R. Loudon, Phys. Rev. A {\bf 55}, 1347 (1997). 

\bibitem{Berini} I. Breukelaar and P. Berini, J. Opt. Soc. Am. A {\bf 23}, 1971 (2006); I. Breukelaar, R. Charbonneau and P. Berini, J. App. Phys. {\bf 100}, 043104 (2006).

\bibitem{ft1} This convention will be used repeatedly throughout.

\bibitem{omegatrans} The delta-transform $\delta(\omega-\omega')\delta(k_{y,i}-k_{y,i}')\delta(q_i-q_i')=(k_i/k_{x,i}v_G)\delta(k_{x,i}-k_{x,i}')\delta(k_{y,i}-k_{y,i}')\delta(q_i-q_i')$ is used to evaluate the orthogonality condition, with $v_G=\partial\omega/\partial k_i$ as the group velocity of the excitation. In addition, the relation $\cos a x +y \sin a x=(r+1)^{-1}[r e^{i ax}+e^{-i ax}]$ is used, where $r=(1-iy)/(1+iy)$.

\bibitem{Blow}  K. J. Blow, R. Loudon, S. J. D. Phoenix and T. J. Shepherd, Phys. Rev. A {\bf 42}, 4102 (1990).

\bibitem{QMequiv} This is analogous to matching the wavefunctions and their derivatives at the interface. For more details, see D. Griffiths and C. A. Steinke, Am. J. Phys. {\bf 69}, 137 (2001).

\bibitem{Marcuv} L. B. Felsen and N. Marcuvitz, {\sl Radiation and Scattering of Waves}, Wiley, New Jersey (1994).

\bibitem{SA} M. Abramowitz and I. A. Stegun (eds.), {\sl Handbook of Mathematical Functions}, Dover, New York (1965).

\bibitem{ST} B. E. A. Saleh and M. C. Teich, {\sl Fundamentals of Photonics}, 2$^{\rm nd}$ Ed., Wiley, New Jersey (2007).

\bibitem{NC} M. A. Nielsen and I. L. Chuang, {\sl Quantum Computation and Quantum Information}, Cambridge University Press, Cambridge (2000).

\bibitem{KLM} E. Knill, R. Laflamme and G. Milburn, Nature {\bf 409}, 46 (2001).

\bibitem{OM} R. A. Campos, B. E. A. Saleh and M. C. Teich, Phys. Rev. A {\bf 40}, 1371 (1989); Z. Y. Ou and L. Mandel, Am. J. Phys. {\bf 57}, 66 (1989); K. S. Mascarenhas, Am. J. Phys. {\bf 59}, 1150 (1991).

\bibitem{CM} C. K. Carniglia and L. Mandel, Phys. Rev. D {\bf 3}, 280 (1971).

\bibitem{Boyd} R. Boyd, {\sl Nonlinear Optics}, 3$^{\rm rd}$ Ed., Academic Press, London (2008).

\bibitem{Pendry} J. B. Pendry, L. Martin-Moreno and F. J. Garcia-Vidal, Science, {\bf 305}, 847 (2004); A. P. Hibbins, B. R. Evans and J. R. Sambles, Science {\bf 308}, 670 (2005); A. P. Hibbins, E. Hendry, M. J. Lockyear and J. R. Sambles, Opt. Express {\bf 16}, 20441 (2008); E. Hendry, A. P. Hibbins and J. R. Sambles, Phys. Rev. B {\bf 78}, 235426 (2008); S. Collin, C. Sauvan, C. Billaudeau, F. Pardo, J. C. Rodier, J. L. Pelouard and P. Lalanne, Phys. Rev. B {\bf 79}, 165405 (2009).

\bibitem{WasserShan} D. Wasserman, E. A. Shaner and J. G. Cederberg, Appl. Phys. Lett. {\bf 90}, 191102 (2007); E. A. Shaner, J. G. Cederberg and D. Wasserman, Appl. Phys. Lett. {\bf 91}, 181110 (2007).

\bibitem{Brown} C. V. Brown, G. G. Wells, M. I. Newton and G. McHale, Nat. Photon. {\bf 3}, 403 (2009).

\bibitem{Ambati} M. Ambati, D. A. Genov, R. F. Oulton and Xiang Zhang, IEEE J. Sel. Top. Quant. Elec. {\bf 14}, 1395 (2008).

\bibitem{Ballester} D. Ballester, M. S. Tame and M. S. Kim, {\it in preparation} (2010).

\bibitem{idealmeas} For a description of such a process and the implications of non-ideal detection, see Ref.~\cite{TSPP,DSPP}.

\bibitem{HOM} C. K. Hong, Z. Y. Ou, and L. Mandel, Phys. Rev. Lett. {\bf 59}, 2044 (1987).

\bibitem{Ghosh} R. Ghosh, C. K. Hong, Z. Y. Ou and L. Mandel, Phys. Rev. A {\bf 34}, 3962 (1986); R. Ghosh and L. Mandel, Phys. Rev. Lett. {\bf 59}, 1903 (1987).

\end{thebibliography}
\end{document}